\documentclass[traditabstract]{aa} 

\usepackage{graphicx}
\usepackage{txfonts}
\usepackage[english]{babel}  
\usepackage{epsfig}

\begin{document}

\title{Magnesium in the atmosphere of the planet  
HD\,209458\,b~: \\Observations of the thermosphere-exosphere transition region}

\titlerunning{Magnesium in the atmosphere of the planet HD\,209458\,b}

   \author{A. Vidal-Madjar\inst{1}
   \and C.~M. Huitson\inst{2}
   \and V. Bourrier\inst{1}
   \and J.-M. D\'esert\inst{3,4,1}
   \and G.~Ballester\inst{5}
   \and A.~Lecavelier~des~Etangs\inst{1} 
   \and D.~K.~Sing\inst{2}
   \and D. Ehrenreich\inst{6}
   \and R. Ferlet\inst{1} 
   \and G. H\'ebrard\inst{1,7}
   \and J.~C. McConnell\inst{8}
 }

   \institute{
      Institut d'Astrophysique de Paris, UMR7095 CNRS, Universit\'e Pierre \& Marie
   Curie, 98bis, boulevard Arago, 75014 Paris, France, \email{alfred@iap.fr}
          \and
    School of Physics, University of Exeter, Exeter, EX4 4QL, UK
              \and
CASA, Department of Astrophysical and Planetary Sciences, University of Colorado, Boulder, CO 80309, USA
              \and
Division of Geological and Planetary Sciences, California Institute of Technology, MC 170-25 1200, E. California Blvd., Pasadena, CA 91125, USA
          \and
   University of Arizona, USA
         \and
     Observatoire de Gen\`eve, Universit\'e de Gen\`eve, 51 Chemin
     des Maillettes, 1290 Sauverny, Switzerland
            \and
   Observatoire de Haute-Provence, CNRS/OAMP, 04870
   Saint-Michel-l'Observatoire, France
            \and
   Department of Earth and Space Science and Engineering, York University 4700 Keele street, Toronto, ON M3J1P3, Canada
   }

   \date{Received TBC; accepted TBC}

  \abstract
   {     

The planet HD\,209458\,b is one of the most well studied hot-Jupiter exoplanets. The
upper atmosphere of this planet has been observed through ultraviolet/optical transit observations with
H\,{\sc i} observation of the exosphere revealing atmospheric escape. At lower altitudes just below the
thermosphere, detailed observations of the Na\,{\sc i} absorption line has revealed an atmospheric
thermal inversion. This thermal structure is rising toward high temperatures at high altitudes, as
predicted by models of the thermosphere, and could reach $\sim$10\,000 K at the exobase level. Here, we
report new near ultraviolet Hubble Space Telescope\,/\,Space Telescope Imaging Spectrograpgh (HST/STIS) observations of atmospheric absorptions during the planetary
transit of HD\,209458\,b. 

We report absorption in atomic magnesium (Mg\,{\sc i}), while no signal has
been detected in the lines of singly ionized magnesium (Mg\,{\sc ii}). We measure the 
Mg\,{\sc i} atmospheric
absorption to be 6.2$\pm$2.9\% in the velocity range from $-62$ to $-19$\,km/s. The detection of atomic
magnesium in the planetary upper atmosphere at a distance of several planetary radii gives a first
view into the transition region between the thermosphere and the exobase, where
atmospheric escape takes place. We estimate the electronic densities needed to compensate for the
photo-ionization by dielectronic recombination of Mg+ to be in the range of 
10$^8$$–-$10$^9$\,cm$^{-3}$. Our
finding is in excellent agreement with model predictions at altitudes of several planetary radii.

We observe Mg\,{\sc i} atoms escaping the planet, with a maximum radial velocity (in the stellar rest
frame) of $-60$\,km/s. Because magnesium is much heavier than hydrogen, the escape of this
species confirms previous studies that the planet's atmosphere is undergoing hydrodynamic escape.
We compare our observations to a numerical model that takes the stellar radiation
pressure on the Mg\,{\sc i} atoms into account. We find that the Mg\,{\sc i} atoms must be present at up to 
$\sim$$7.5$~planetari radii altitude and estimate an Mg\,{\sc i} escape rate 
of $\sim$$3\times$$10^7$~g s$^{-1}$. Compared to previous
evaluations of the escape rate of H\,{\sc i} atoms, this evaluation is compatible with a magnesium 
abundance roughly solar. A hint of absorption, detected at low level of
significance, during the post-transit observations, could be interpreted as a Mg\,{\sc i} cometary-like tail.
If true, the estimate of the absorption by Mg\,{\sc i} would be increased to a higher value of about
8.8$\pm$2.1\%.
}

   {}
   {}
   {}
   {}


  \maketitle

%


\section{Introduction}

The first detection of an exoplanet atmosphere was accomplished by detecting sodium in the
transiting hot-Jupiter HD\,209458\,b (Charbonneau et al.\ 2002). A few other atomic species have been
identified in the evaporating upper atmosphere of this planet as well, including hydrogen,
oxygen and carbon (Vidal-Madjar et al.\ 2003, 2004). An extended hydrogen envelope of this planet
was also suggested by Hubble Space Telescope / Advanced Camera for Surveys (HST/ACS) observations (Ehrenreich et al.\ 2008), although these data were
not significantly conclusive. The atmospheric escape mechanism has been identified to be an
hydrodynamic ``blow-off'' (Vidal-Madjar et al.\ 2004). While Ben Jaffel et al.\ (2007) confirmed the
detection of an extended upper atmosphere, the author debated the escape mechanism (see
also Vidal-Madjar et al.\ 2008 and Ben Jaffel 2008). However, more recent observations obtained
with the Hubble Space Telescope / Cosmic Origins Spectrograph (HST/COS) have independently confirmed the nature of the hydrodynamic escape mechanism
for this planet (Linsky et al.\ 2010).

Interestingly, complementary observations of the exoplanet's atmosphere have been secured at
different wavelengths, hence probing different altitudes. Ballester et al.\ (2007) reported detection
of H\,{\sc i} from recombination via the Balmer jump, while Rayleigh scattering by H$_2$ molecules has also
been shown to be a likely explanation for the observed increase of planetary radii toward 
near ultraviolet (NUV)
wavelengths (Sing et al.\ 2008a, 2008b; Lecavelier des Etangs et al.\ 2008b).

Complementary observations have also probed deeper in the atmosphere of HD\,209458\,b with
signatures of molecular species detected using transit spectrophotometry (Knutson et al.\ 2007,
Sing et al.\ 2008a) or dayside spectrum (Swain et al.\ 2009). 
From this spectrum, broad band signatures were interpreted as due to the
presence of water vapor (Barman 2007); upper-limits on TiO/VO abundances at high altitude were
estimated (D\'esert et al.\ 2008), and the thermosphere has also been revealed through a detailed
analysis of the Na\,{\sc i} line profile (Vidal-Madjar et al.\ 2011a, 2011b). 
Near-infrared observations revealed the presence of molecules deeper 
in this planet's atmosphere with detections of CO (Snellen et al.\ 2010) and H$_2$O (Deming et al.\ 2013).
Hot-Jupiter orbit so close to their parents stars that they are exposed to intense extreme ultraviolet (EUV) irradiation
and strong stellar winds, which can shape their atmospheres (Lecavelier des Etangs et al.\ 2004).
The high temperatures cause the
atmosphere to escape rapidly, implying that the upper thermosphere is cooled primarily by adiabatic
expansion (Yelle et al.\ 2004). One of the hottest exoplanets known is the hot-Jupiter, WASP-12b
(Hebb et al.\ 2009). Fossati et al.\ (2010) and Haswell et al.\ (2012) have detected the escaping upper
atmosphere of this planet using HST/COS spectra obtained in the NUV. In particular,
their observations focused on the extra absorption observed in the Mg\,{\sc ii} resonance line cores. These
authors interpreted their results in the framework of hydrodynamical escape, similarly to the case of
HD\,209458\,b. In general, UV transit observations have the potential to reveal the mass-loss rates of
an exoplanet atmosphere (e.g., Ehrenreich \& D\'esert 2011, Bourrier \& Lecavelier des Etangs 2013).

Here, we report new observations of HD\,209458\,b's extrasolar atmosphere in the NUV. This
spectral domain is particularly rich in terms of the detection possibilities of high altitudes species,
which have strong absorption signatures at these wavelengths. These observations were completed
under the Hubble Space Telescope (HST) program (ID\#11576, PI J-M. D\'esert) completed with the
Space Telescope Imaging Spectrograph (STIS) instrument between July and December 2010.
Observing in the UV offers a great advantage, because of the strong opacity of atoms and ions in
that spectral range and the ease in identifying the species that causes the observed absorption, in sharp
contrast with the interleaved molecular bands in the near-IR. We first report the observations and
provide details on the instrumental systematics that affect these observations in Section~2, while we
present the data analysis in Section~3. We discuss our findings and draw conclusions from our
observations in the two last sections of the paper.

\begin{figure}
   \centering
   \includegraphics[width=\columnwidth]{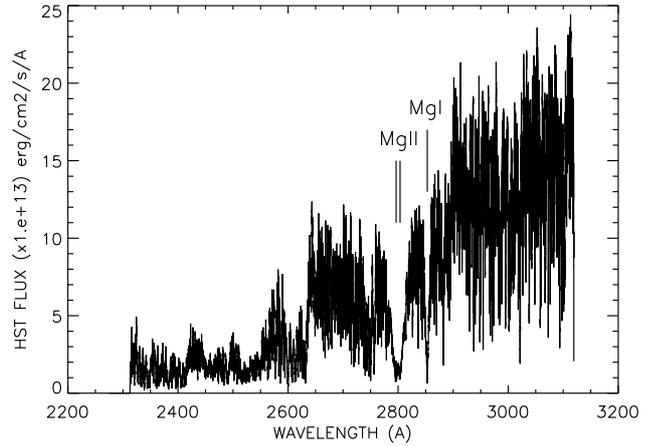}
   \caption{The spectrum of HD\,209458  
   over the whole observed spectral range.
   Many spectral signatures are seen in the stellar spectrum which includes
   the Mg\,{\sc ii} doublet near 2\,800\AA , its core emissions due to the
   stellar chromosphere, and the strong Mg\,{\sc i} line near
   2\,850\AA .}
     \label{spectrum}
\end{figure}

\section{Observations and corrections}
\label{obs_section}

The observations were conducted 
using the Echelle E230M STIS grating, which provides a resolving
power of $\lambda/\Delta\lambda=30\,000$. All observations were
completed with the NUV-Multi-Anode Microchannel Array 
(NUV-MAMA) detector and a 0.2\arcsec x\,0.2\arcsec
entrance aperture, providing about 2 pixels per resolution
element.

Three transits of HD\,209458\,b were observed; each used a
succession of five consecutive HST orbits.
Each HST orbit is built upon a similar observing sequence,
producing ten successive exposures of 200 seconds each
(except for the first orbits, which contain only nine exposures due to 
the time needed to the acquisition). 
Table~\ref{observations} lists the observations log. 

To compute the planetary orbital phase at the time of observation, 
we assumed a planet's orbital period of 3.52474859~days and
a central transit time of T$_0$=53344.768245 (MJD)
(Knutson et~al.\ 2007). The transits starts and ends at
$-0.018$ and $+0.018$ planetary orbital phases, respectively. The mid-orbital phases
shown in Table~\ref{observations} give the
positions of the five HST orbits during each transit: observations of 
orbits\#1 and \#2 are always before 
the transits, orbits\#3 and \#4 always during the transits, 
and orbit\#5 always after the transits.

The spectra were extracted with the standard \textsc{calstis} pipeline
(version 2.32), which includes localization of the orders, optimal
order extraction, wavelength calibration, corrections of
flat-field, etc. 
An example of this spectral extraction is shown in Fig.~\ref{spectrum}
with wavelengths between $\sim$~2\,300\AA\
and 3\,100\AA\ at the nominal spectral resolving power, which is at about 
10~km/s resolution. These spectra contain large amounts of transitions, which gives a  
noisy appearance.

\begin{table}
\centering
\caption{Times are for the mid-exposure of each orbit.}
\begin{tabular}{cccc}
 \hline
 Observing time   & Transit~~HST-Orbit  & Heliocen. & Planetary \\
   &   & Corr. & orbital \\
  (UT in 2010) & ~~\#~~~~~~~~~~\# & (km/s)  & phase \\
 \hline
Jul. 18 11:51:29   &  ~~1~~~~~~~~~~1  & --18.01 & --~0.0492\\
Jul. 18 13:23:21   &  ~~1~~~~~~~~~~2  & --17.99 & --~0.0311\\
Jul. 18 14:59:16   &  ~~1~~~~~~~~~~3  & --17.98 & --~0.0122\\
Jul. 18 16:35:10   &  ~~1~~~~~~~~~~4  & --17.96 & ~~~0.0067\\
Jul. 18 18:11:04   &  ~~1~~~~~~~~~~5  & --17.94 & ~~~0.0256\\
 \hline
Sep. 09 08:26:17   &  ~~2~~~~~~~~~~1  & 3.13 & --~0.0527\\
Sep. 09 09:59:27   &  ~~2~~~~~~~~~~2  & 3.16 & --~0.0343\\
Sep. 09 11:35:21   &  ~~2~~~~~~~~~~3  & 3.19 & --~0.0155\\
Sep. 09 13:11:15   &  ~~2~~~~~~~~~~4  & 3.22 & ~~~0.0034\\
Sep. 09 14:47:09   &  ~~2~~~~~~~~~~5  & 3.25 & ~~~0.0223\\
 \hline
Dec. 06 11:43:12   &  ~~3~~~~~~~~~~1  & 26.44 & --~0.0491\\
Dec. 06 13:15:40   &  ~~3~~~~~~~~~~2  & 26.43 & --~0.0309\\
Dec. 06 14:51:30   &  ~~3~~~~~~~~~~3  & 26.43 & --~0.0120\\
Dec. 06 16:27:21   &  ~~3~~~~~~~~~~4  & 26.43 & ~~~0.0069\\
Dec. 06 18:03:11   &  ~~3~~~~~~~~~~5  & 26.43 & ~~~0.0257\\
 \hline
\end{tabular}
\label{observations}
\end{table}

\subsection{STIS echelle orders overlaps}

Observations made with the E230M echelle spectrograph give access to several orders, 
covering the whole spectral domain but with some overlap at the extremity of these orders. 
In the overlapping spectral regions, we averaged the measurements obtained from the
two different overlapping orders. At each orders edge, a few pixels (three consecutive ones, corresponding to about 15~km/s velocity domains or less than 0.15~\AA\ wide) were found to be unreliable. 
For these pixels, we simply interpolated the spectrum measured at shorter and longer wavelengths,
keeping track of these overlapping wavelengths. For instance, they are listed for Transit\#2 in Table~\ref{oders}.
Because all three transits are not observed at the same epoch ({\it i.e.} at different heliocentric corrections, see Table~\ref{observations}), the three transits evaluated radial velocities in the heliocentric rest frame can present relative shifts 
up to the instrument spectral resolution of 10~km/s.

\begin{table}
\centering
\caption{Table of E230M orders overlapping wavelengths for the Transit \#2 (in \AA).}
\begin{tabular}{cccc}
 \hline
    2339.10   &    2366.01   &    2393.55   &    2421.73  \\
    2450.59   &    2480.14   &    2510.40   &    2541.42  \\
    2573.21   &    2605.81   &    2639.24   &    2673.54  \\
    2708.74   &    2744.87   &    2781.98   &    2820.11  \\
    2859.30   &    2899.58   &    2941.02   &    2983.66  \\
    3027.55   &    3072.74   &                &               \\
 \hline
\end{tabular}
\label{oders}
\end{table}

As a result, one should be careful to interpret 
any detected signature at the spectral orders edges, where an interpolation is done
(covering $\pm$0.40 \AA ).
When broad band spectral regions are considered (of about 200\AA ), we find the potential perturbations are on the order of 0.1\%\ of the studied signal and thus 
are considered negligible.
In contrast, it is important to check that the positions of the order 
overlaps does not coincide with the studied domain for narrow spectral bands. In the present study, this is never the case.

\subsection{Orbital time series observations and correction of the STIS thermal ``breathing'' effect}
\label{correction of the STIS thermal ``breathing'' effect}

For each transit observation, we discard data obtained during the first HST orbit, 
because systematic trends in the first orbit are found to be always significantly 
worse than in subsequent orbits;
 this is likely due to the fact that the HST must thermally relax into its new pointing 
position during this time. The procedure of discarding the first HST orbit of each transit 
is usually performed for HST observations of exoplanets' transits ({\it i.e.}, 
Sing et al.\ 2011; Huitson et al.\ 2012). For each HST orbit containing 10 sub-exposures, 
we also discard the first sub-exposures, which are always found to have an anomalous low flux. 

\begin{figure}[htb!]
   \centering
   \includegraphics[width=9cm]{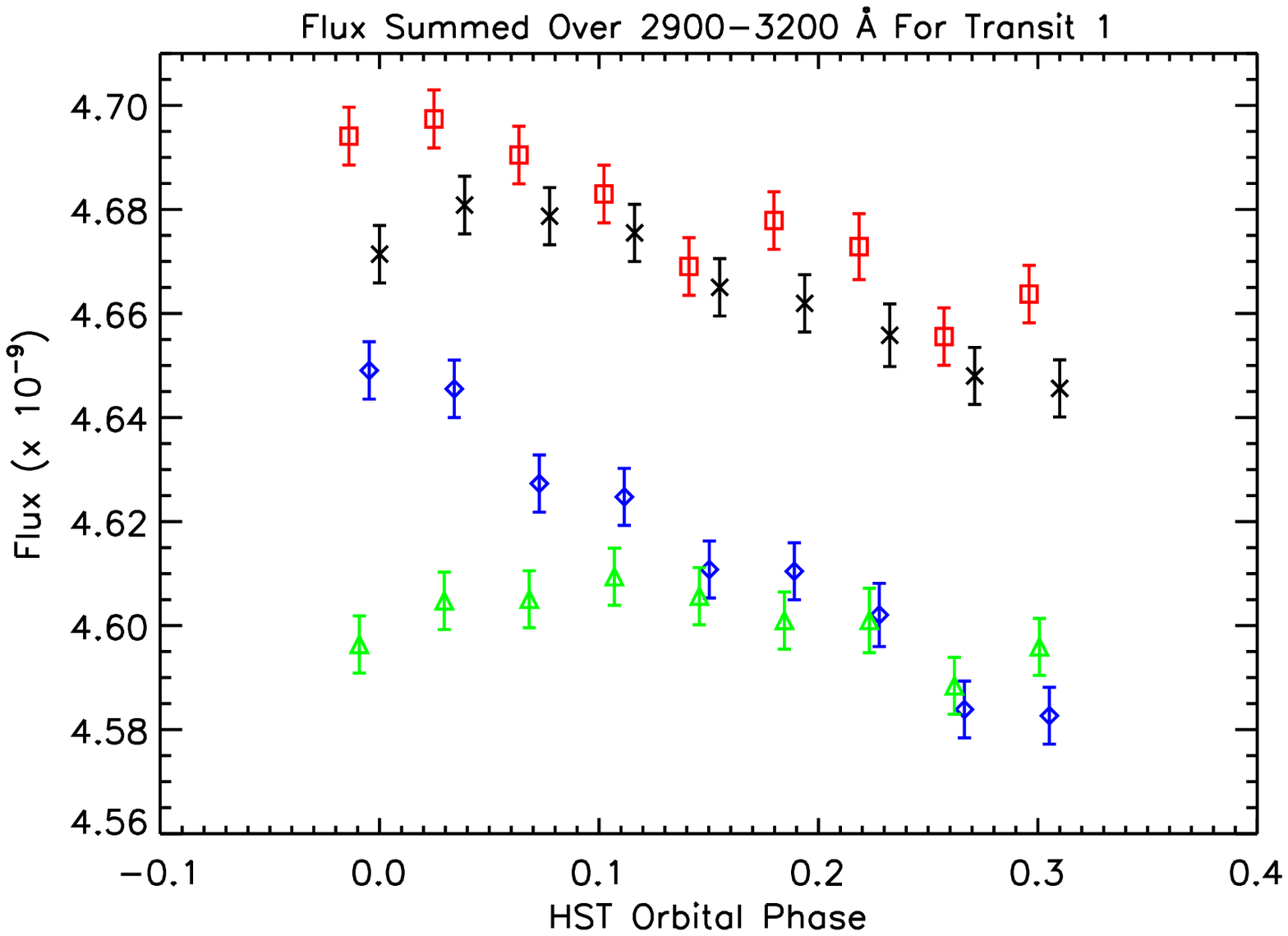}
   \includegraphics[width=9cm]{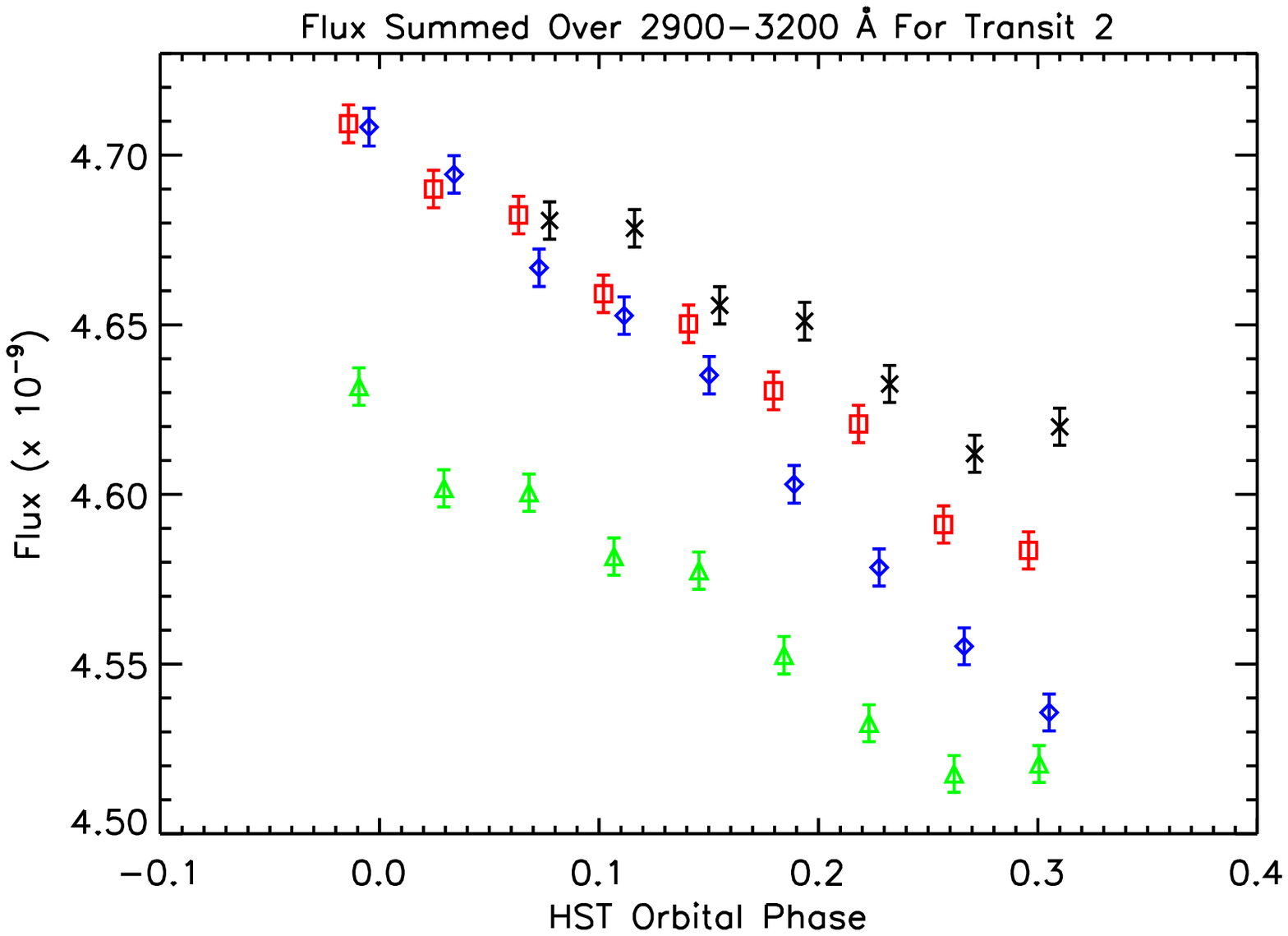}
   \includegraphics[width=9cm]{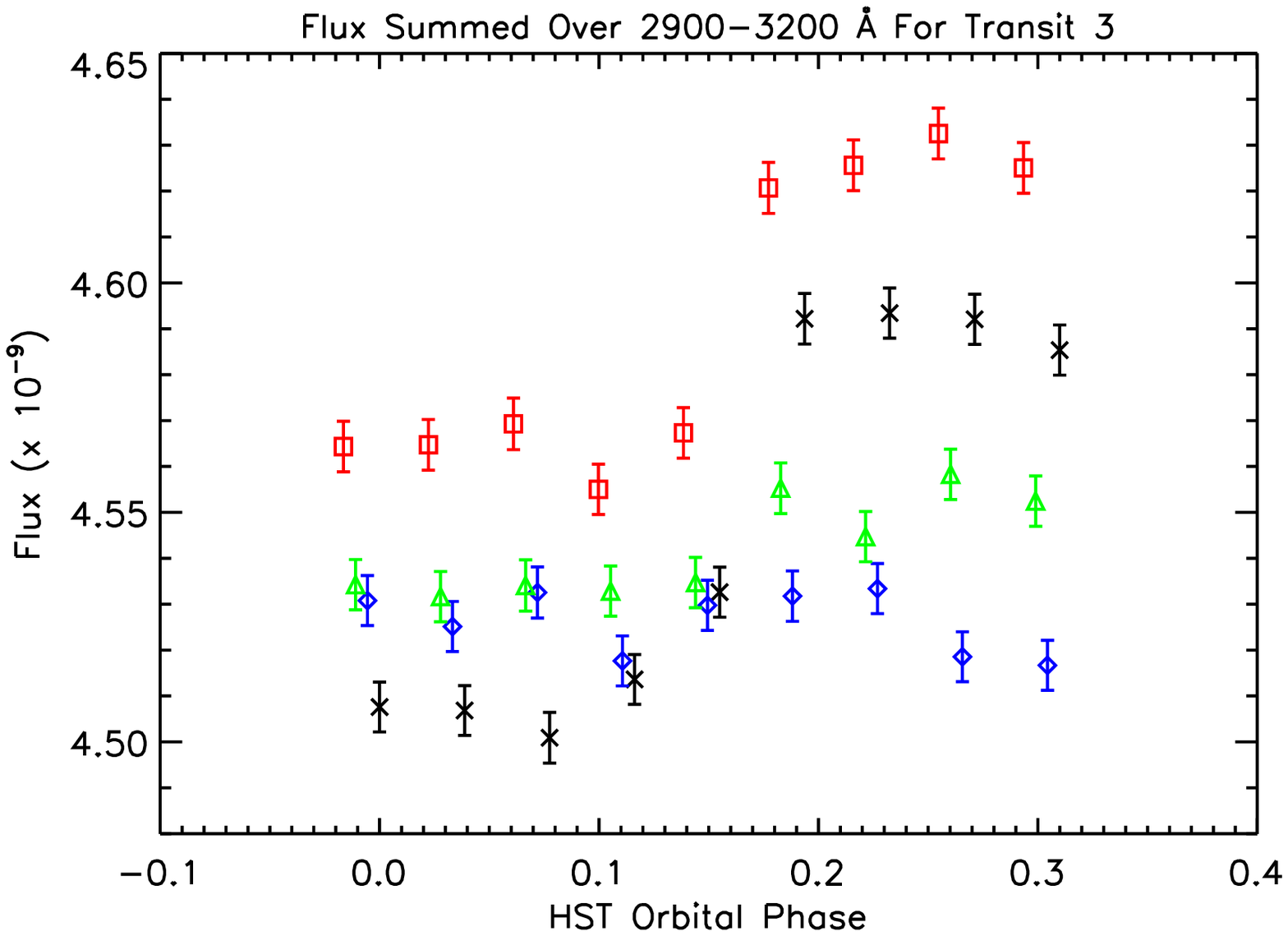}
   \caption{The total HST flux of each sub-exposure measured over the 2900--3100\AA\
   spectral domain as a function of the HST orbital phase (in erg.cm$^{\rm -2}$.s$^{\rm -1}$).
   {\bf Upper plot.} Transit\#1. During this transit, successive 
   HST orbits are shown (orbit\#2, black crosses, orbit\#3, blue diamonds, orbit\#4, 
   green triangles and orbit\#5, red squares). 
   {\bf Middle plot.} Same as upper plot for transit\#2.
   {\bf Lower plot.} Same as upper plot for transit\#3.}
   \label{Breathing}
\end{figure}

The time series observations during each orbit are then extracted by adding the flux over a given spectral domain for each of the considered sub-exposures. As an example, this is shown in Fig.~\ref{Breathing}, where the total flux is summed over the 2\,900--3\,100~\AA\ wavelength range 
for each sub-exposure of each orbit of each visit as a function of the HST orbital phase.  
Clear trends are seen that are repeated in all orbits. The correlation of the flux with 
the HST orbital phase is typical of the HST/STIS thermal ``breathing'' effect as the HST is heated 
and cooled during its orbit, causing focus variations. 

The shape and amplitude of these systematic variations can change 
between visits of the same target, as has been found in the STIS optical transit observations of HD 209458 (Brown et al.\ 2001; 
Charbonneau et al.\ 2002; Sing et al.\ 2008a,b) and HD 189733 (Sing et al.\ 2011; Huitson et al.\ 2012).   
Extensive experience with the optical STIS data over the last decade further indicates that the orbit-to-orbit variations 
within a single visit are both stable and highly repeatable outside of the first orbit, which displays a different trend 
(see Brown et al.\ 2001 and Charbonneau et al.\ 2002).  Because of its repeatability, this effect can be corrected.

It can be seen in Fig.~\ref{Breathing} that all orbits in transits~\#1 and \#2 
show similar trends, although the observations obtained
during the planet's transit (the in-transit orbit: orbit~\#4) and during ingress (orbit~\#3) 
have lower baseline flux levels compared to the corresponding out-of-transit orbits; this 
is  due to the planetary transit itself. 

To correct for the systematic trends introduced by the HST ``breathing'' effect, 
the transit depths were fitted simultaneously with a 4$^{\mathrm{th}}$ order polynomial 
function of HST orbital phase, which is referred to hereafter as the ``breathing'' correction. 
This correction has proved to be very successful for removing systematic trends 
in past observations using STIS ({\it e.g.}, Sing et al.\ 2008a; Ehrenreich et al.\ 2012; 
Bourrier et al.\ 2013). 
We investigated whether higher order terms were justified 
by calculating the Bayesian Information Criterion (BIC), which  
is defined as $\chi^2+k \ln n$ for a normal distribution, where $k$ is the number of free parameters and $n$ 
is the number of data points. 
It was found that higher order terms that are 
greater than 4$^{\mathrm{th}}$ order were not justified.

\begin{figure}[htb!]
   \centering
   \includegraphics[width=9cm]{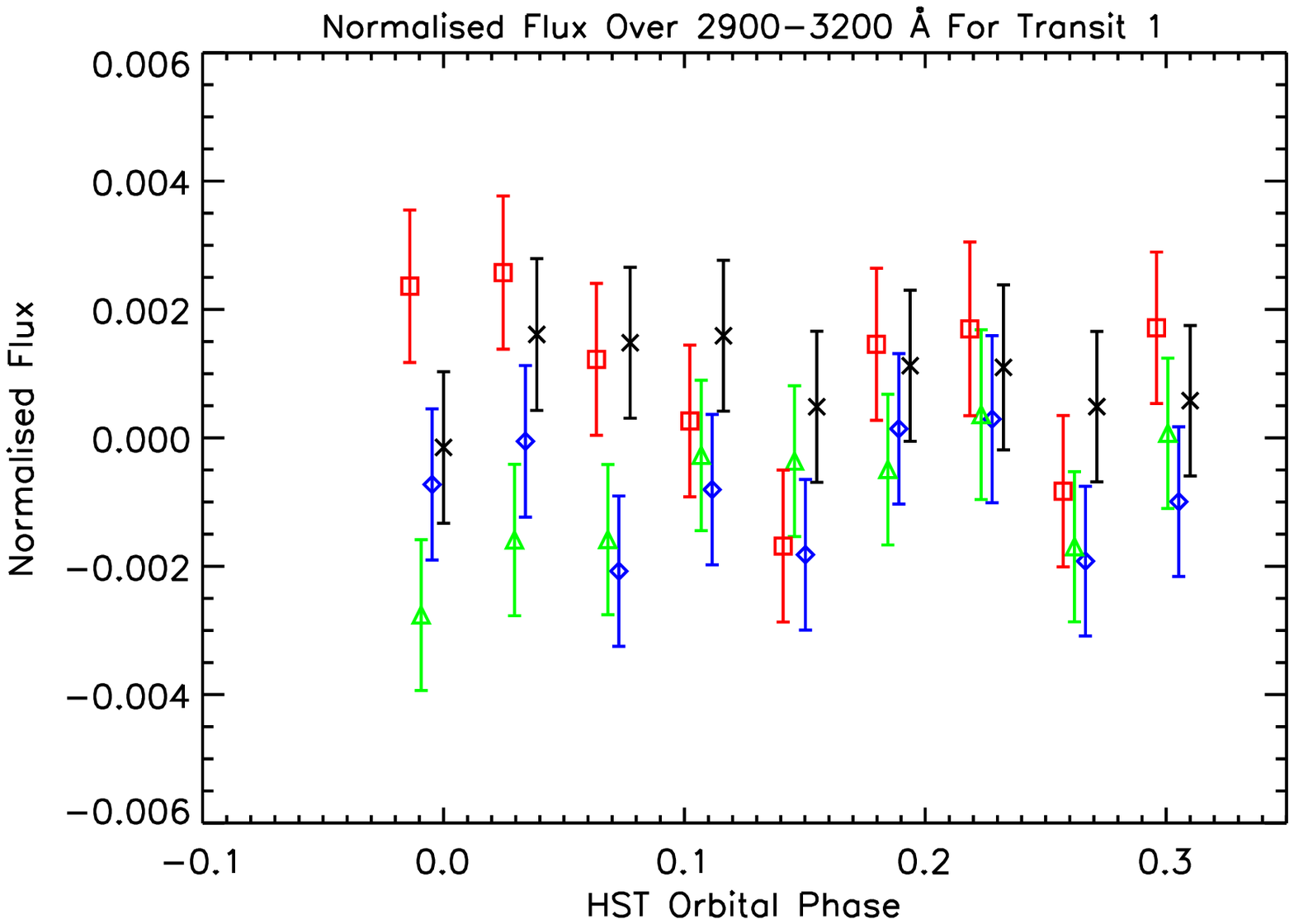}
   \includegraphics[width=9cm]{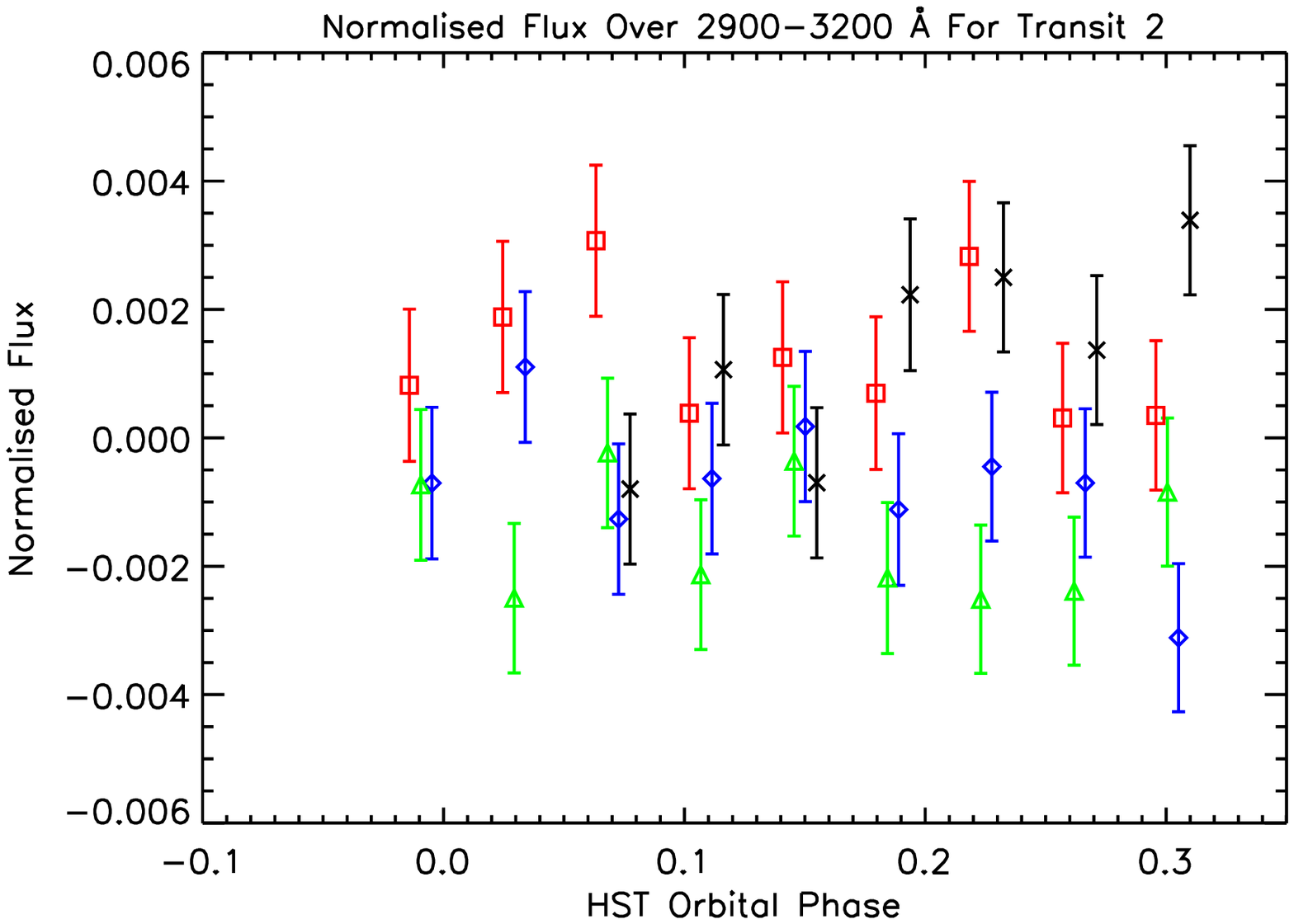}
   \includegraphics[width=9cm]{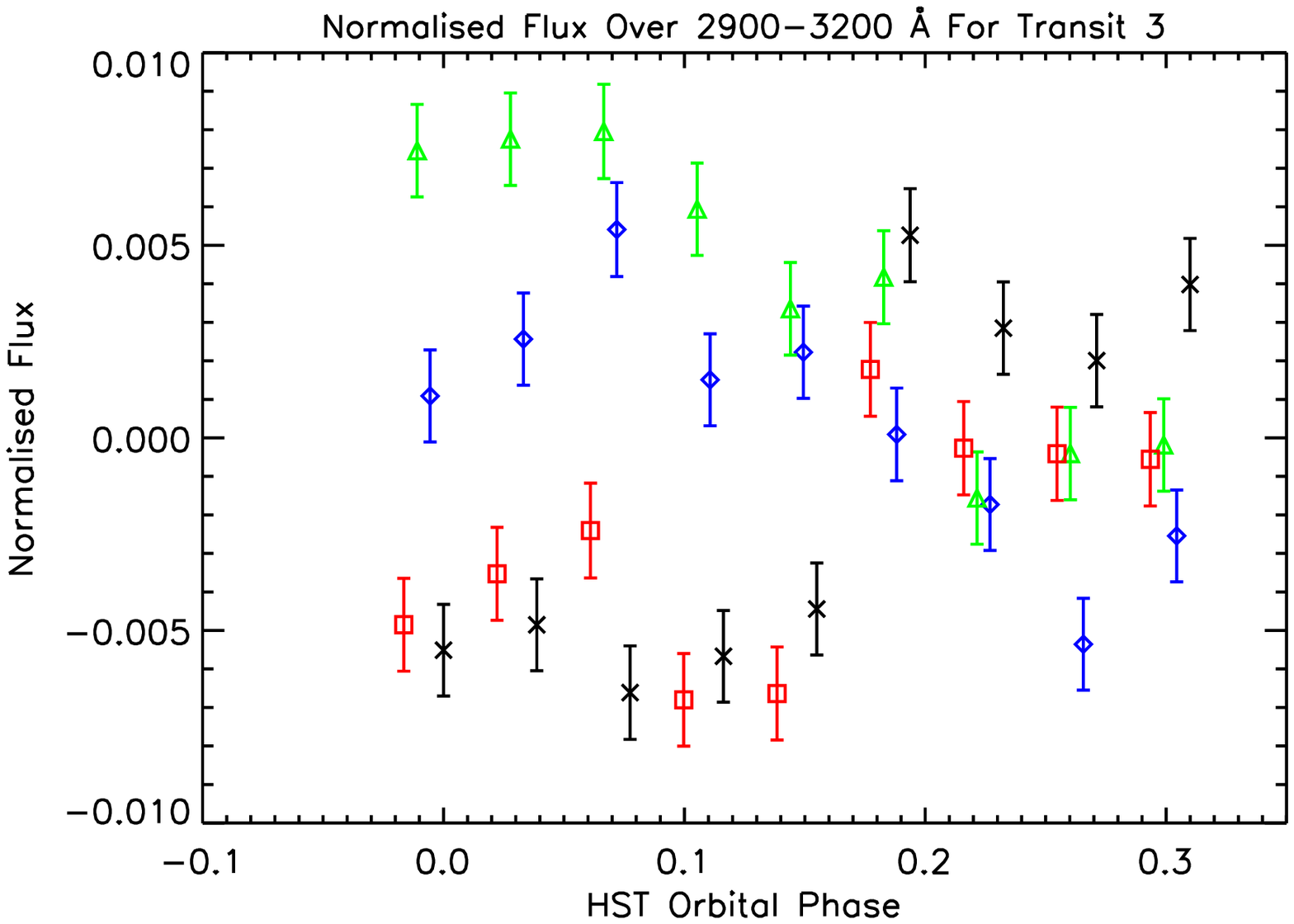}
   \caption{Same as Fig.~\ref{Breathing} for the fluxes of each sub-exposure 
   normalized by the total flux that is measured during the transit and corrected  
   for all trends as discussed (see text). 
   {\bf Upper plot.} Transit\#1. The overall fluctuation after correction 
   is $\pm$~0.2\%.
   {\bf Middle plot.} Same as upper plot for transit\#2. Overall 
   fluctuation~:  $\pm$~0.3\%.
   {\bf Lower plot.} Same as upper plot for transit\#3. Note here the
   different ordinate scale showing the relatively more perturbed transit\#3. 
   For that transit, the overall fluctuation is $\pm$~0.7\%.}
   \label{Breathing_corr}
\end{figure}

In the case of transit~\#3, the 4$^{\mathrm{th}}$ order polynomial correction 
still leaves a $\pm$~0.7\% residual fluctuation (Fig.\ref{Breathing_corr}). 
When a better accuracy is required ({\it i.e.}, for transit 
``absorption depths'', or $ADs$ of the order of 1.5\% or less), 
we ignore transit~\#3, which brings more noise than information. 
In contrast, for larger $ADs$,
transit~\#3 is included in the analysis.

\subsection{Baseline of the light curves and stellar activity} 
\label{sec_visit_long_flux_corr}

The variation in flux between orbits, as seen in transit~\#3, could be due to stellar variability.
In the UV, where spectral lines can be very sensitive to the stellar activity,
variations on the order of $\pm$1.5~\% in the UV are not unexpected. 
For example, some chromospheric lines can display variations 
of $\sim 2-3$~\% per hour as in the case of the Sun (see {\it e.g.} Vidal-Madjar 1975, Lemaire 1984). 

In the case of the Mg\,{\sc i} and Mg\,{\sc ii} spectral line studies presented here, it is known that the Mg\,{\sc ii} line cores are formed from the higher parts of the photosphere to the upper part of the chromospheric plateau (Lemaire \& Gouttebroze 1983), while the Mg\,{\sc i} lines are formed in the upper part of the solar photosphere to the low chromosphere (Briand \& Lemaire 1994). This means that the Mg\,{\sc ii} line cores are more sensitive to stellar activity (in the case of a G2V star) than the Mg\,{\sc i} line core. Because no significant
variations are seen in the core of the Mg\,{\sc ii} lines during
transits (even during the third one, see Sect.~\ref{The MgII spectral absorption signature}) at a level of 2\% at most, this means that we should expect even less perturbations due to stellar activity in the case of the Mg\,{\sc i} line core study. Because the observed Mg\,{\sc i} variations
are above 6\% (Sect.~~\ref{The MgI spectral absorption signature}), we consider that variations
due to stellar activity of the solar-like star HD\,209458 are small enough to be properly corrected from now on, as was done with previous studies.

This comment also concerns the question of the stellar disk homogeneity. Such star-spot-like signatures, already seen in more active stars, could have some influence on the transit depth/shape and thus on the evaluations of the fitted parameters. The star HD\,209458 which is of solar type and is quiet at the time of our observations since the Mg\,{\sc ii} perturbation, is $\sim$~2\%, any stellar disk inhomogeneity should affect less the Mg\,{\sc i} (see spatial observations of the solar disk, {\it e.g.} Bonnet \& Blamont 1968; Lemaire \& Gouttebroze 1983). Again, Mg\,{\sc i} show absorption signatures at the 6\% level; these evaluations should not be affected significantly by possible stellar, spatial, or temporal activity. 
 
In previous datasets ({\it e.g.}, Sing et al.\ 2008a; 
Sing et al.\ 2011; Huitson et al.\ 2012), visit-long flux variations 
have been corrected by fitting 
a linear slope as a function of planetary phase for the baseline of the light curves
during the entire visits. A simple linear slope over timescales of an HST-transit 
is found to be sufficient because stellar variability cycles 
have a typical duration of the order of several days. 
We also investigated the possibility of better fits to the data by light curves
using higher degrees of the polynomial for the flux baseline. 
We found that the BIC does 
not improve significantly when using the 2$^{\mathrm{nd}}$ degree polynomial 
for the baseline when binning the spectrum into 200~\AA\ bins. For narrower spectral bands, the effect is even smaller. 
We therefore decided to use a linear function for the flux baseline 
in the fit to the light curves. This further shows that the stellar activity during our observations is low, as a perturbing stellar spot signature seen along the path of the planetary disk in front of the star would have led toward the selection of a higher degree polynomial.

\subsection{Fit of the light curves and de-trending model}
\label{The de-trending model}

For all spectral bands used in the analysis presented in the following sections, 
we applied the breathing correction (a $4^{\mathrm{th}}$ order function of the HST phase) 
and a linear baseline for the out-of transit light curve 
to consider the visit-long stellar flux variations (linear function of the planetary phase).
We used the model of Mandel \& Agol (2000) to fit the transit light curve. 
The three transits were fitted simultaneously with the same functional form for the corrections but 
with separate correction parameters for each visit. 
One common value for the $R_P/R_\star$ parameter was used to fit all visits simultaneously. 
The orbital inclination of the system, central transit time, and $a/R_\star$ are fixed to the values 
from Hayek et al.\ (2012).

The de-trending model 
used for each visit is described by
\begin{equation}
F=F_0 * (a \phi + b_1 \phi_{HST} + b_2 \phi_{HST}^2  + b_3 \phi_{HST}^3  + b_4 \phi_{HST}^4 +1),
\label{equation_model_detrend}
\end{equation}
where $F_0$ is the baseline stellar flux level, $a,b_1, b_2, b_3$, and $b_4$ 
are constant fitted parameters, $\phi$ is planetary phase,
and $\phi_{HST}$ is HST orbital phase (for the breathing correction). 
These parameters are found by a fit to the data; 
they are different for each of the three visits. 

Finally, we use the Kurucz (1993) 1D ATLAS 
stellar atmospheric models\footnote[1]{See http://kurucz.harvard.edu} 
and a 3-parameter limb darkening law to correct for the stellar limb darkening of the form:
\begin{equation}
\frac{I(\mu)}{I(1)}=1-c_{2}(1-\mu)-c_{3}(1-\mu^{3/2})-c_{4}(1-\mu^2),
\end{equation}
where $\mu = \cos(\theta)$ and $\theta$ is the angle in radial direction from the disc center
(see details in Sing 2010). 
The models were calculated at a very high resolution, R~=~500\,000, 
to ensure that the limb darkening could be well described for the numerous spectral lines 
in the data.  We used T$_*$~=~6\,000~K, log~g~=~4.5, and metallicity~=~0.0 
as the closest approximation to HD\,209458.
The limb-darkening coefficients 
are fixed when fitting the planet transit light curve.

\subsection{Error estimation}
\label{Errors estimation}

To obtain an estimate of the absorption depth, ($AD$), that is
translatable into a planetary radius, we have to use the complete de-trending 
model and include the limb-darkening correction to the fit of the data,
as described in Sect.~\ref{The de-trending model}. 
However, the error bars produced by \textsc{mpfit} (Markwardt 2009), the code that we used
to perform the 
simultaneous fit of the transit model and systematics correction, 
are an overestimate.
Indeed, \textsc{mpfit} determines error bars using the full 
covariance matrix, which means that it considers 
the effect of each parameter on the other parameters. 
Such error estimates overestimate differential absorption 
depths if a large portion of the systematic trends are common mode. 
This is discussed in more details for a specific case 
(see below Section~\ref{The MgI spectral absorption signature}).

Therefore, 
we used the comparison between in-transit and out-transit 
spectra and calculated the corresponding absorption depth ($AD_{In/Out}$)
to estimate the significance of a possible detected signal. 
By comparing the evaluated flux over any given bandpass, 
over which the stellar flux is averaged in a similar manner 
for orbits~\#2 and \#5 $(F_2+F_5)$ on one hand 
and for orbits~\#3 and \#4 $(F_3+F_4)$ on the other, 
we have access to the relative absorption depth ($AD_{In/Out}$)~:
$$
AD_{In/Out}=1-\frac{(F_3+F_4)}{(F_2+F_5)}
$$
Using the errors provided by the STIS pipeline for each pixel of the corresponding 
observation, we can evaluate the error $E_{In/Out}$ on the
evaluated $AD_{In/Out}$~:
$$
E_{In/Out}=\frac{(F_3+F_4)}{(F_2+F_5)}\times \sqrt{\frac{E_3^2+E_4^2}{(F_3+F_4)^2}  +  \frac{E_2^2+E_5^2}{(F_2+F_5)^2}}.
$$
Because the orbits \#4 are completed during the ingress of each of the three planetary transits, 
the $AD_{In/Out}$ estimate is expected to be lower than the $AD$ obtained through a full 
fit of the light curve (Sect.~\ref{The de-trending model}). Because the in/out ratio
is free from uncertainties introduced by correlation in the parameters of the fit, it 
can be used to obtain estimates of error bars on differential absorption depths and 
detection significance levels.

\section{Data analysis}
\label{data_analysis}

\subsection{Broadband transit absorption depths}
\label{Continuum absorption at 2900--3100}

\begin{figure}[htb]
   \centering
   \includegraphics[width=\columnwidth]{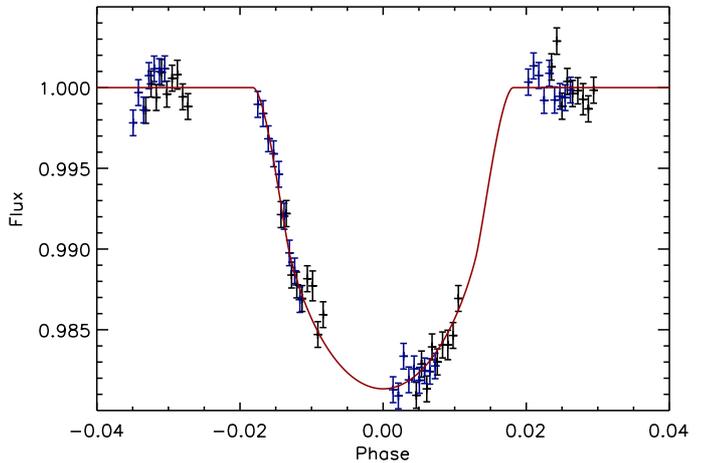}
   \caption[]{
    Transit light curves over the broadband 2\,900--3\,100~\AA\ that
   show the measurements obtained over each individual sub-exposure 
   (black dots for transit\#1, and blue dots for transit\#2) as a function of the 
   planet's orbital phase. The first and last group of points, before and after the transit, 
   around phase -0.34 and +0.25 respectively, correspond to the Out of transit
   orbits~\#2 and \#5, while the two other groups around phase -0.15 and +0.05 
   correspond to the In transits observations that are completed respectively during orbit~\#3
   (during ingress) and orbit~\#4 (deep within the transit).
   The solid red line shows the fitted profile,  
   including limb-darkening and instrument breathing corrections.  
      }
   \label{2900_3100}
\end{figure}

We first looked for the transit signal over broad bands 
in the stellar continuum, which can be compared to
the transit that is previously observed at longer wavelengths in the optical. 
At wavelengths of about 5\,000 \AA , the depth of the transit light curve 
was measured to be $AD$$\approx$($R_P/R_*$)$^2$=1.444\%, where
R$_P$ and R$_*$ are the planetary and stellar radii, respectively 
(Sing et al.\ 2008a, 2009). 
From the variation of $AD$ between 4\,000~\AA\ and 5\,000~\AA\ and assuming that
the absorption at these wavelengths is dominated by Rayleigh scattering 
in the planet atmosphere (see Lecavelier des Etangs et al.\ 2008a), 
the $AD$ at 3\,000~\AA\ is expected to be $AD_{3\,000}$~$\sim$1.484\%.
Using a fit to the data of transits~\#1 and \#2 and the de-trending model as
described in Sect.~\ref{The de-trending model}, 
we measured the planetary radii in 200~\AA\ bins
(Fig.~\ref{2900_3100}).
Here, we excluded the transit~\#3 data because it shows clear 
residual trends as a function of time
(Sect.~\ref{correction of the STIS thermal ``breathing'' effect}).
Table~\ref{coeff} presents the coefficients as evaluated according 
to Eq.~\ref{equation_model_detrend}, and Table~\ref{table_broad_ad} 
shows the $AD$ evaluated for the different spectral domains.

\begin{table}
\centering
  \begin{tabular}{c | c | c }
\hline
Transit~\# & 1 & 2 \\
\hline    
a & 0.045515510 & -0.076254266 \\
$b_1$ & 0.014154763 & -0.14127084 \\
$b_2$ & -0.22566153 & 1.2484295 \\
$b_3$ & 0.22764628 & -7.2927994 \\
$b_4$ & 0.43358347 & 12.768201 \\
\hline
\end{tabular}
\caption{Eq.~\ref{equation_model_detrend} coefficients used in the de-trending 
model of the two transits \#1 and \#2.
}
\label{coeff}
\end{table}

\begin{table}
\centering
  \begin{tabular}{c | c | c | c }
\hline
Start &  End & $AD$ & $AD$ \\
Wavelength & Wavelength & (\%) & Error \\
(\AA) & (\AA) & & (\%) \\
\hline  
       2\,300   &    2\,500   &    1.31   &  0.11 \\
       2\,500   &    2\,700   &    1.55   &  0.07 \\
       2\,700   &    2\,900   &    1.51   &  0.05 \\
       2\,900   &    3\,100   &    1.46   &  0.03 \\       
\hline
\end{tabular}
\caption{Broadband absorption depths ($AD$) measured over various spectral 
domains using the two first transit observations.
}
\label{table_broad_ad}
\end{table} 

The $AD$ measured here are consistent with 
the extrapolation, assuming Rayleigh scattering.
However, our evaluations are not accurate enough to confirm
that the variation of the $AD$ as a function of wavelength follows the
Rayleigh scattering law; {\it a fortiori}, this cannot be 
used to obtain an estimate of the atmospheric temperature.
This limitation is mainly caused by the decrease in the 
stellar flux at short wavelengths, as can be seen
in Table~\ref{table_broad_ad}, where the 
errors clearly increase below 2\,700~\AA . 

\begin{figure}[htb]
   \centering
   \includegraphics[width=\columnwidth]{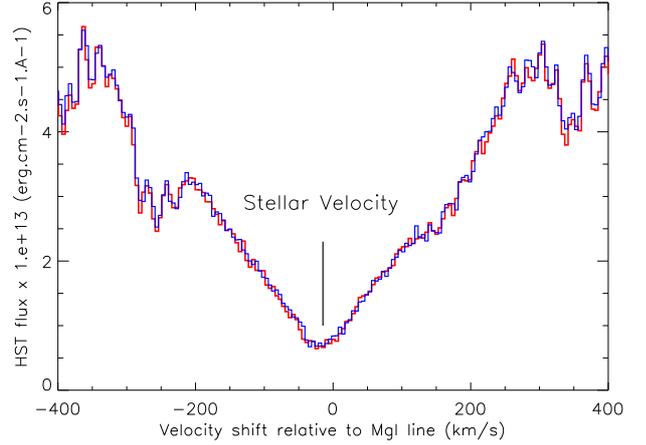}
   \includegraphics[angle=90,width=\columnwidth]{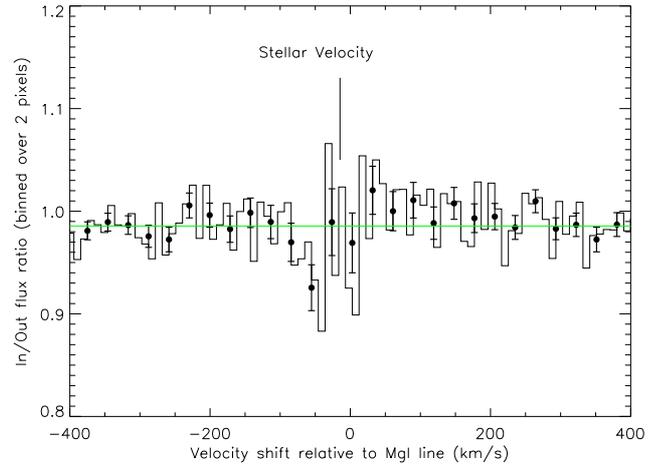}
   \caption{The Mg\,{\sc i} line spectral region.
   {\bf Upper panel.} The spectrum observed during the three transits.
   The average of orbits~\#2 and \#5 (blue line) represent the
   Out of transit spectrum while the average of orbits~\#3 and
   \#4 (red line) represent the In transit spectrum.
   Wavelengths are transformed into velocity shifts relative to the
   Mg\,{\sc i} rest wavelength (individual pixels correspond to about 5~km/s bandwidth).
   {\bf Lower panel.} The In/Out 
   ratio of the two spectra (orbits \#3 and \#4 divided by
   orbits~\#2 and \#5). The horizontal green line at the $AD$=1.444\%\ level
   corresponds to the Sing~et al.\ (2008a) absorption depth 
   evaluation at 5\,000~\AA . The histogram shows the ratio 
   rebinned over one resolution element which is
   over two instrument pixels (about 10 km/s).
   The black dots give the ratios rebinned by 6 pixels 
   with corresponding error bars, showing a single 
   feature detected at more than 2-$\sigma$ over a wide range of wavelengths. 
}
   \label{MgI_spetrum}
\end{figure}

\subsection{The Mg\,{\sc i} spectral absorption signature}
\label{The MgI spectral absorption signature}

Magnesium is an abundant species, and its neutral form 
presents a unique and strong spectral line, 
Mg\,{\sc i} at 2\,852.9641\AA\ (Fig.~\ref{MgI_spetrum}). 
This deep line of about 500~km/s in width allows the 
measurement of the star radial velocity, which agrees 
with previous estimates (V$_{Star}$=-14.7~km/s; Kang et al.\ 2011).

In Fig.~\ref{MgI_spetrum}, the Out of the transit spectrum is
the average of orbits \#2 and \#5, while the In transit one is
the average of orbits \#3 and \#4. In both cases, 
these averages are completed over the three planet transit observations. 
The inclusion or exclusion of the third transit does not significantly alter the results.
Around the Mg\,{\sc i} line center, the In spectrum is seen to be
lower than the Out spectrum (upper panel of Fig.~\ref{MgI_spetrum}).
The In/Out ratio shows an excess absorption over several dozens 
of km/s in the blue side of the line 
(lower panel of Fig.~\ref{MgI_spetrum}). 
To evaluate precisely the absorption depth, we then use the fitting
model described above.

\subsubsection{The de-trending model study}
\label{Standard approach}

\begin{figure}[htb]
   \centering
   \includegraphics[width=\columnwidth]{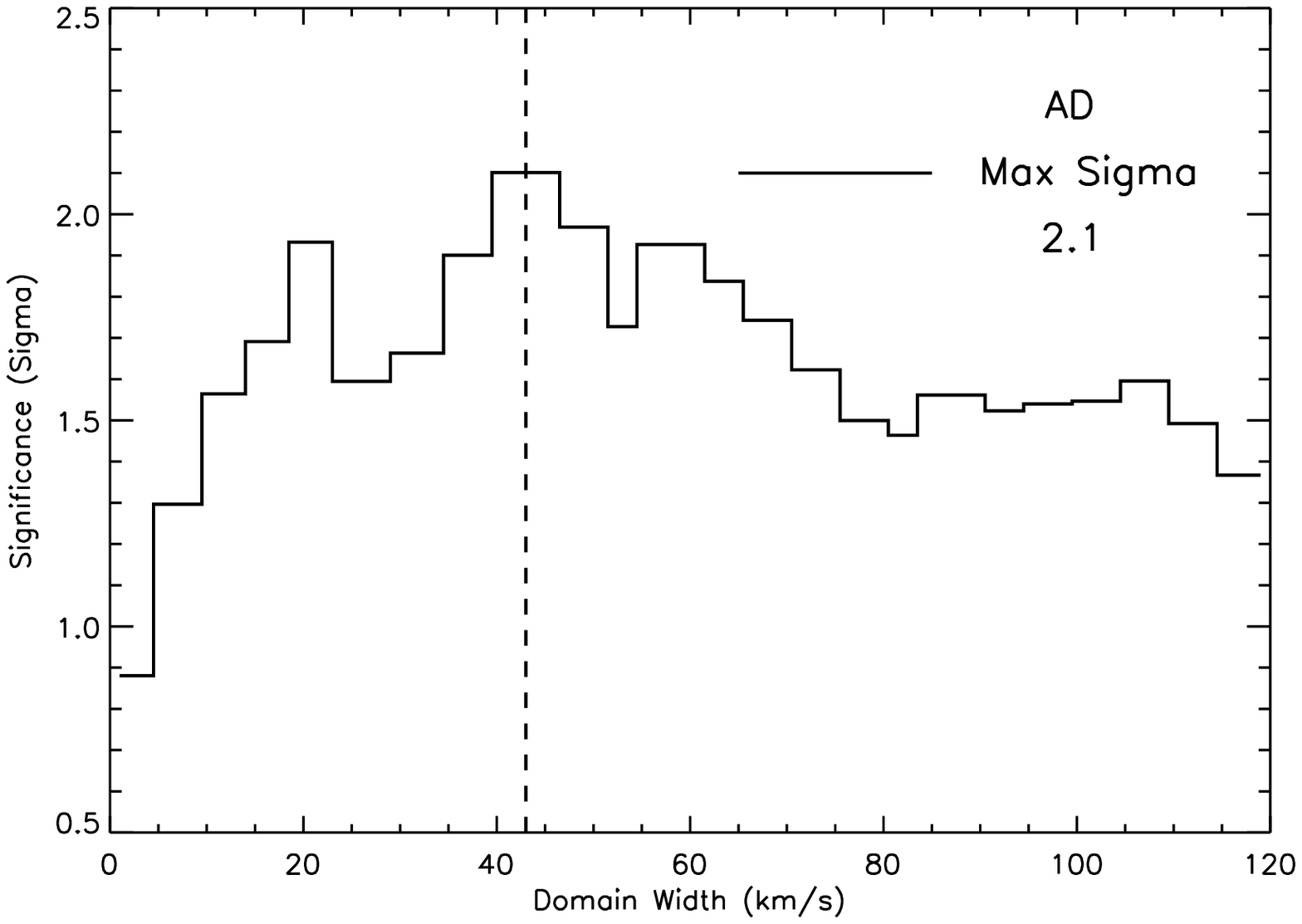}
   \includegraphics[width=\columnwidth]{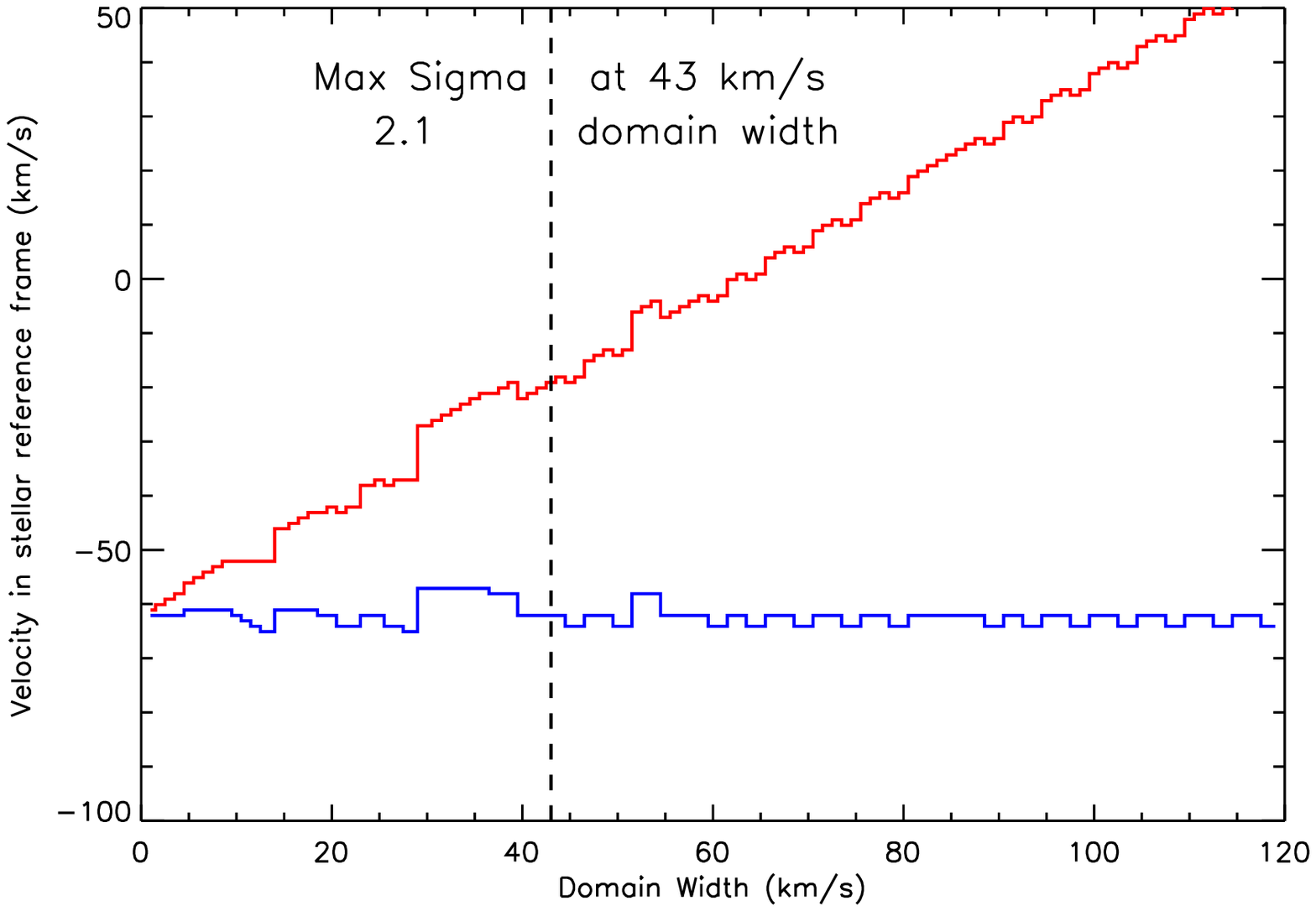}
   \caption{
   {\bf Upper panel.} Plot of the maximum detection level
   of the absorption depth 
   in the Mg\,{\sc i} line as a function of the band width. 
   The maximum detection level is reached at the $\sigma$-ratio of 2.1 
   for a band width of 43~km/s$\pm$3 km/s ($AD$=6.2$\pm$2.9\%).
   {\bf Lower panel.} The velocity limits, V$_{\rm min}$ (blue line) and  
   V$_{\rm max}$ (red line), of the  
   band domains where the maximum of the significance level is reached 
   as a function of the band width. 
   V$_{\rm min}$ is found to be roughly constant at
   about -62~km/s (stellar reference frame). 
   The position of the
   highest maximum signal-to-noise ratio at 43~km/s is shown by 
   a vertical dashed line. 
}
   \label{MgI_bandwidth}
\end{figure}

\begin{figure}[htb]
   \centering
     \includegraphics[width=\columnwidth]{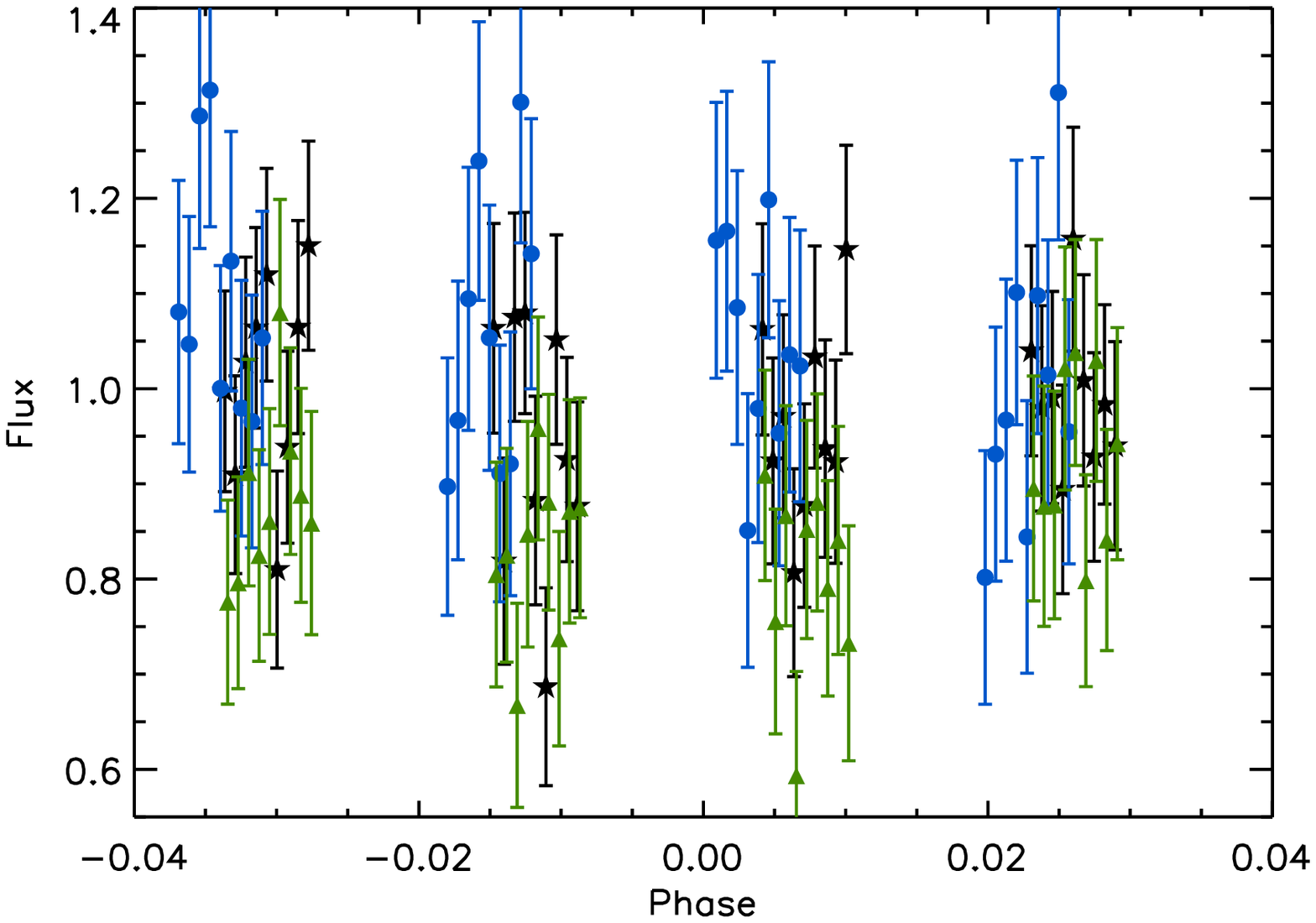}
     \includegraphics[width=\columnwidth]{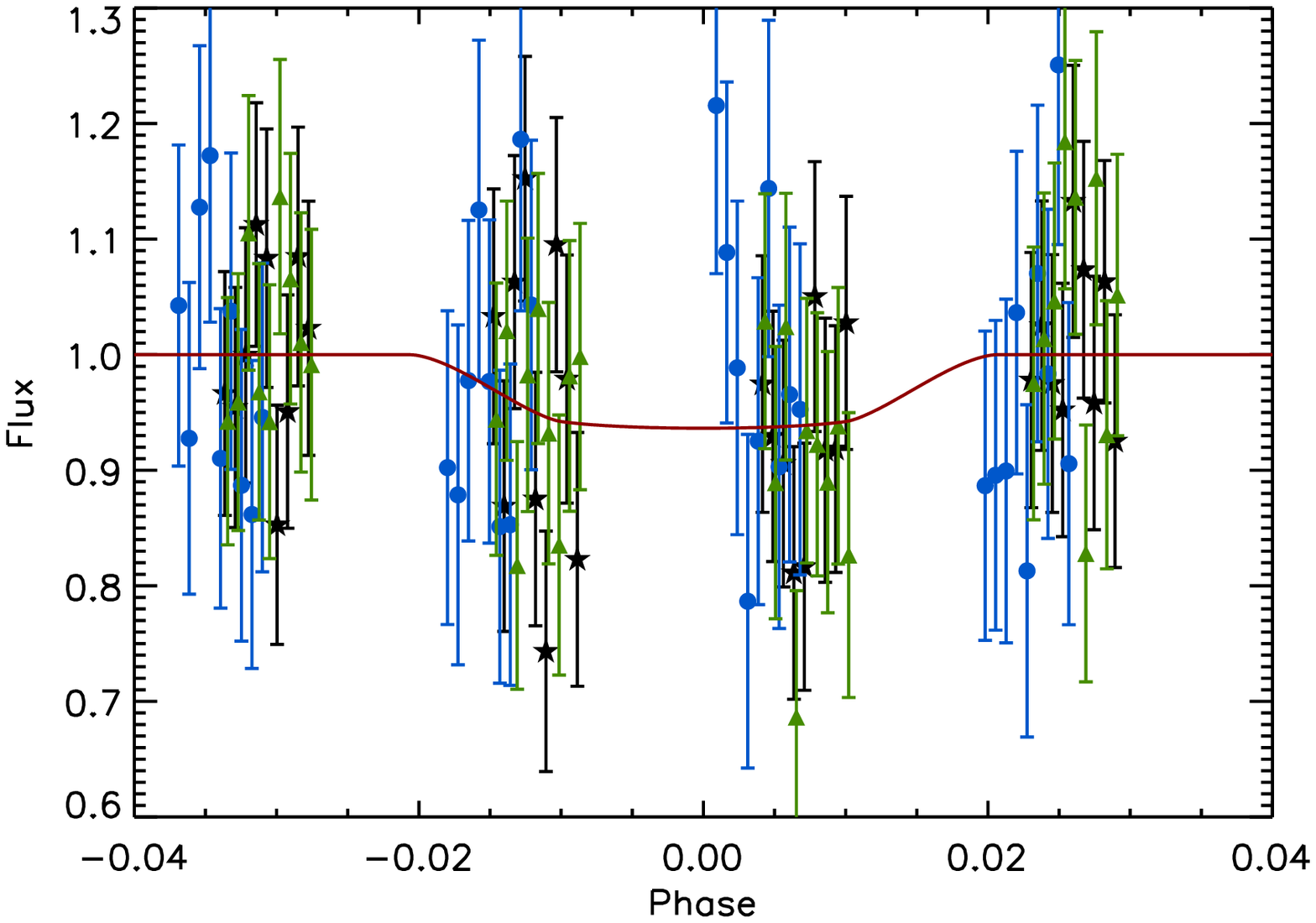}
   \caption[]{
   The transit light curve measured in 
   the -62 to -19 km/s spectral domain. The out-of-transit reference corresponds
   to the measurements in the orbits~\#2 (before transit) 
   and \#5 (after transit). {\bf Upper panel.} Raw observations. 
   {\bf Lower panel.} The limb-darkening and all trend correction that are included along 
   with the final
   global fit (red solid line). 
   All subexposures (black dots for transit~\#1, blue dots for transit~\#2 and green dots for transit~\#3) are shown. The evaluated $AD$ that corresponds to the spectral domain is equal to 6.2$\pm$2.9\%. Note how flattened the limb-darkening looks: the reason being that the larger Mg\,{\sc i} absorption gives a significantly larger planetary radius, which acts to smooth out the limb-darkening profile.
   }
   \label{MgI_-62_-19}
\end{figure}

To determine the width and velocity of the Mg\,{\sc i}
absorption signature, as suspected from the ratio plot 
(lower panel of Fig.~\ref{MgI_spetrum}), we fitted the data using the de-trending 
model over the three transits and evaluated 
the $AD$ for a variable spectral band width. We found the 
spectral position for which the ratio $\sigma=AD/E_{AD}$ 
is the largest, where $E_{AD}$ is the uncertainty on the measured absorption depth. 
As seen in Fig.~\ref{MgI_bandwidth}, 
the strongest signature is found for a band width of 43~km/s 
extending from -62 to -19~km/s. We note that 
the blue edge of the absorption domain is sharp, as shown
by the constant value of the minimum velocity of the band that provides
the largest $\sigma$-ratio for the absorption (V$_{\rm min}$ around 
-62~km/s) and independent of the band width. 

Since the evaluated absorption is over $\sim$6\% and
much larger than the nearby continuum $AD^{Cont.}$ levels ($\sim$1.5\%), 
the transit~\#3 can be used in the estimates obtained above, 
because the data of this transit does not introduce more than $\pm$0.7\% fluctuations. 

The atmospheric Mg\,{\sc i} transit signal over narrow bandpasses, which is
in excess to the transit seen over large bandpasses and where Mg\,{\sc i} 
does not absorb the stellar light, 
$AD^{Excess}$, can be evaluated by comparing the total absorption 
depth ($AD^{Total}$) in the considered spectral region ({\it i.e.} near 2\,853 \AA ) 
to the absorption signature in the nearby continuum ($AD^{Cont.}$).
The $AD^{Cont.}$ is  
evaluated over two symmetric broad spectral domains ($\sim~$2~\AA\ wide) on both sides 
of the Mg\,{\sc i} line (from -2\,500 to -500~km/s in the blue side
of the line and from +500 to +2\,500~km/s in the red side).
This leads to an average nearby
stellar continuum absorption, $AD^{Cont.}$=1.4$\pm$0.1\%. 

To estimate the excess absorption in the spectral line due to the atmosphere only, 
this value must be subtracted to the total absorption measurement. 
From the estimated absorption 
depth $AD^{Total}$=6.2$\pm$2.9\% and by using the relation,
$$
AD^{Excess}=1-\frac{(1-AD^{Total})}{(1-AD^{Cont.})},
$$
we evaluate an Mg\,{\sc i} absorption excess of $AD^{Excess}$=4.9$\pm$2.9\% .

The whole de-trending model which includes the limb darkening correction, 
is adjusted over the observations for the spectral domain by providing the highest 
detection level for the absorption (from -62 to -19~km/s; Fig.\ref{MgI_-62_-19}).
We can however question the error bars estimates made via the fitting process 
because of the large number of parameters that are introduced 
and simultaneously fitted (breathing, light curve baseline, 
and the limb darkening corrections). 
To test these errors, we perturbed each parameter in the fit 
by $\pm 1$~$\sigma$ and found that the $AD^{Excess}$ value changed by less than 0.1\%, while the difference 
in $AD^{Total}$ changed by 0.9\% depending on the de-trending 
parameters used. 
We also found that a very similar spectrum
as the one shown in Fig.\ref{MgI_43kms_Orb2} is produced by not fitting for trends at all.  
This seems to indicate that photon noise clearly dominates systematic 
noise in our small bins. Additionally, we found that the limb darkening correction 
made very little difference to the measured $AD$ values; this is likely due to 
the large size of the planet's effective radius and
the flatness of limb darkening at this wavelength 
(and as seen by comparison of Fig.~\ref{2900_3100} and ~\ref{MgI_-62_-19}).
The difference between the $AD$ obtained with the limb darkening correction 
using the model coefficients and without the correction 
entirely changed the $AD^{Excess}$ value by only 0.5\%.

\subsubsection{The $AD_{In/Out}$ study}

To further investigate the systematic effects, we evaluated 
the same absorption signatures by conducting a similar study 
at this wavelength range
and using the $AD_{In/Out}$ evaluations (Sect.~\ref{Errors estimation}).
For the same spectral domain as above (from -62 to -19~km/s), 
the In/Out estimates lead to
a total absorption depth of $AD_{In/Out}^{Total}$=7.5$\pm$1.6\% . 
The average stellar continuum absorption is $AD_{In/Out}^{Cont.}$=1.2$\pm$0.1\% . 
Thus, the Mg\,{\sc i} absorption excess is $AD_{In/Out}^{Excess}$=6.4$\pm$1.6\% .
This corresponds to a detection level of the atmospheric Mg\,{\sc i} of $\sigma_{In/Out}$=4.0.

Here, the detection level of the absorption signature of atmospheric 
Mg\,{\sc i} is higher. This is not because the absorption depth value has 
changed (from 6.2 to 7.5\%), but occurs when the error evaluation 
is significantly lower (from 2.9 to 1.6\%).  
We can explain this result by the fact that the 
systematics are strongly tight together 
between the different orbits datasets and that their effect is 
almost washed out when comparisons are made between different orbits. They are 
even more washed out when relative $AD^{Excess}$ evaluations 
are made using the procedure of comparison between in-transit 
and out-transit data.

Using a planetary radius given by (R$_P$/R$_*$)$^2$=1.444\%, 
an absorption of 7.5\%\ corresponds to an opaque magnesium atmosphere that extends up to about 2.3~R$_P$,
an altitude level in the thermosphere/exosphere region where the
blow off process starts to be effective
(see {\it e.g.}, Koskinen 2013a, 2013b).

\subsubsection{Significance and summary of the Mg\,{\sc i} spectral absorption signature}
\label{Final MgI spectral signature}

We calculated the false-positive probability (FPP) to find a Mg\,{\sc i} absorption excess 
that was at least as significant as the one detected in the data but caused by noise only. 
This requires defining {\it a priori} the kind of signatures that would be considered 
{\it a posteriori} as an atmospheric absorption feature. 
In particular, we need to define the wavelength range (and the corresponding velocity range) 
around the Mg\,{\sc i} line center, where any absorption could be considered 
as a plausible absorption due to magnesium in the planet environment. 
Observationally, the highest velocity ever detected in an exoplanet atmosphere is -230~km/s 
in the case of H\,{\sc i} in the blue wing of the Lyman-alpha line 
(Lecavelier des Etangs et al.\ 2012). 
No absorption signature was also ever detected at more than +110~km/s in the red wing 
of a stellar line during transit observations (Bourrier et al.\ 2013). 
On the theoretical side, numerical simulations show that the radial velocity 
of a magnesium atom pushed by radiation pressure cannot exceed -230~km/s 
when it transits in front of the stellar disk of HD\,209458. We thus considered a \emph{conservative} velocity 
range of [-300; 150]~km/s, for which any absorption may have been 
interpreted as possibly caused by atmospheric magnesium. 
This conservative velocity range yields a FPP of 13\%. 
Nonetheless, all the observed properties of the signature are well explained 
by our simple dynamical model (Sect.~\ref{Dynamics of escaping magnesium atoms}), 
which can explain absorption only within the velocity range [-230; 0]~km/s. 
This velocity range yields a corresponding FPP of 6\%. If we consider
only the features with properties (velocity, strength, absorption profile) 
which can be reproduced by this model with plausible physical quantities (for ionization, escape rate, etc.) as a positive detection, 
the corresponding FPP would be even lower than 6\% but this is difficult to quantify.

There is also an apparent absorption feature around 0~km/s in Fig.~\ref{MgI_spetrum}. 
This feature is spread over at most 4 pixels (two resolution elements). 
We also calculated its FPP and found that this 
feature has about 75\% chance to be spurious and only due to statistical noise in the data.

\subsection{The Mg\,{\sc ii} spectral absorption signature}
\label{The MgII spectral absorption signature}

With the detection of the Mg\,{\sc i} signature 
(Sect.~\ref{The MgI spectral absorption signature}), the acknowledgement
that ionized species have already been detected 
in the HD\,209458\,b planetary atmosphere (Vidal-Madjar et~al. 2004; 
Linsky et~al. 2010), and the acknowledgement that Mg\,{\sc ii} was possibly detected 
in another extrasolar planet atmosphere (Fossati et~al. 2010; 
Haswell et~al. 2012), we thought there was a good prospect for the detection of 
ionized magnesium 
through the strong doublet at 2\,803.5305~\AA\ (Mg\,{\sc ii}-h) and 2\,796.3518~\AA\ 
(Mg\,{\sc ii}-k).
The presence of a doublet is a clear advantage, because this allows us to 
check both lines for absorption signatures. 
With the doublet lines spectral separation translated into radial velocities on the
order of 770~km/s, any signal due to ionized magnesium with no more than a
$\pm$~400~km/s radial velocity separation benefits from two independent simultaneous observations. 

\begin{figure}[htb]
   \centering
   \includegraphics[width=\columnwidth]{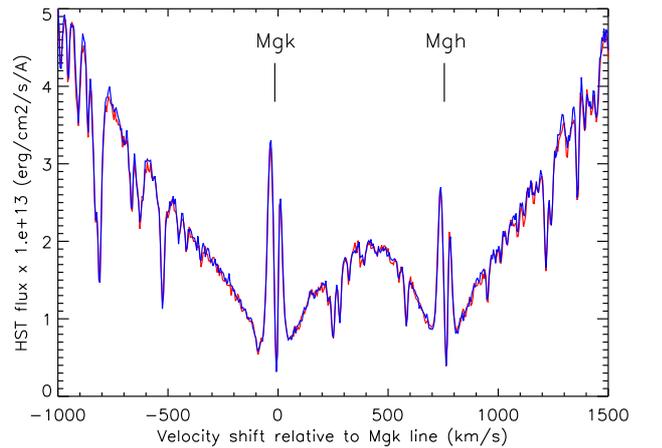}
   \caption[]{The Mg\,{\sc ii} lines spectral region 
   (Mg\,{\sc ii}-k line at 2\,796.35\AA\ 
   and Mg\,{\sc ii}-h line at 2\,803.53\AA).
   The spectrum is observed during the two 
   first transits.
   The average of orbits~\#2 and \#5 (blue line) represents
    the Out of transit spectrum, while the average of 
   orbits~\#3 and
   \#4 (red line) represents the In transit spectrum.
   Wavelengths are transformed into velocity shifts relative to the
   Mg\,{\sc ii}-k line (individual pixels correspond to about 5~km/s bandwidth).
}
   \label{MgII_spetrum}
\end{figure}

\begin{figure}[htb]
  \centering \includegraphics[width=\columnwidth]{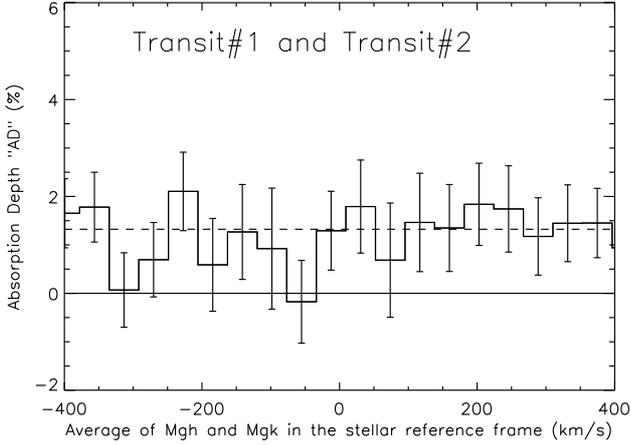} 
   \caption{Plot of the absorption depth in 43~km/s band widths (histogram), as
   measured around the Mg\,{\sc ii} doublet at $\sim$2\,800\AA\ by using orbit~\#2 
   and \#5 as the 
   out-of-transit reference. 
   The absorption depth is averaged 
   over the two Mg\,{\sc ii} lines. 
   The absorption depth in the stellar continuum (dashed horizontal line) 
   is measured using the two sides of the Mg\,{\sc ii} doublet 
   (from -3\,000 to -1\,000~km/s and from +1\,000 to +3\,000~km/s).
}
   \label{MgII_43kms}
\end{figure}

Figure~\ref{MgII_spetrum} shows the spectral region of the Mg\,{\sc ii} doublet. 
A self-reversed emission can be seen in the core of each line, 
which is due to the stellar chromosphere. 
Additional absorption at the center of that chromospheric emission 
is mainly due to the interstellar absorption, which is slightly 
redshifted relative to the star reference frame.

We searched for Mg\,{\sc ii} absorptions using 43~km/s wide spectral domains 
(Fig.~\ref{MgII_43kms}). We did not detect any Mg\,{\sc ii} absorption down to 1\%\ $AD$ level.
We found that the average absorption depth in all of the 43~km/s bins corresponds exactly 
to the absorption depth measured in the nearby stellar continuum.
We also searched over broader and narrower domains and found no significant 
absorption signature. Finally, we also searched for Mg\,{\sc ii} signatures independently 
within each of the three transits observations without any positive result.

The non-detection of excess absorption in the Mg\,{\sc ii} lines (with upper limit of about 
1\% in $AD$) is in strong constrast with the Mg\,{\sc i} detection with $AD$$\sim$6.2--7.5\% 
(Sect.~\ref{The MgI spectral absorption signature}). 
This shows that most of the magnesium must be neutral up to at
least 2~R$_P$ distance from the planet center. 

\section{Discussion}

\subsection{Mg\,{\sc ii} non-detection} 

The non-detection of Mg\,{\sc ii} absorption from the planetary atmosphere cannot 
be interpreted by strong interstellar extinction as in the case of the WASP-12 system 
(Fossati et~al. 2010; Haswell et~al. 2012). 
Indeed, stellar Mg\,{\sc ii} chromospheric emissions 
in the core of the two lines are clearly seen in the case of HD\,209458, showing a low interstellar absorption. 
In light of the Fossati et~al.\ (2010) early ingress detection of the planetary 
transits in the Mg\,{\sc ii} lines, we searched for pre- or post-transit absorption signatures 
in the Mg\,{\sc ii} light curve. We did not find any significant signal before 
or after the transit in neither the data of the orbits~\#2 nor of the orbits~\#5. 
As a result, we conclude that no Mg\,{\sc ii} absorption is detected with an upper limit
of 1\%\ on the absorption depth.

\subsection{Search for a cometary-like tail signature}
\label{Search for a cometary like tail signature}

Because a transit of a cometary-like tail can be observed \emph{after} the planetary transit, 
which is possibly during orbit~\#5, we have compared orbit~\#2 (pre-transit) 
and orbit~\#5 (post-transit). 
For this we have evaluated the absorption depth due to the putative gaseous tail 
($AD_{{In/Out},{Tail}}$) by measuring the difference between the flux as measured before and 
after the transit: 
$$
AD_{In/Out,{Tail}}=\frac{F_2-F_5}{F_2}.
$$

\begin{figure}[htb]
  \centering \includegraphics[width=\columnwidth]{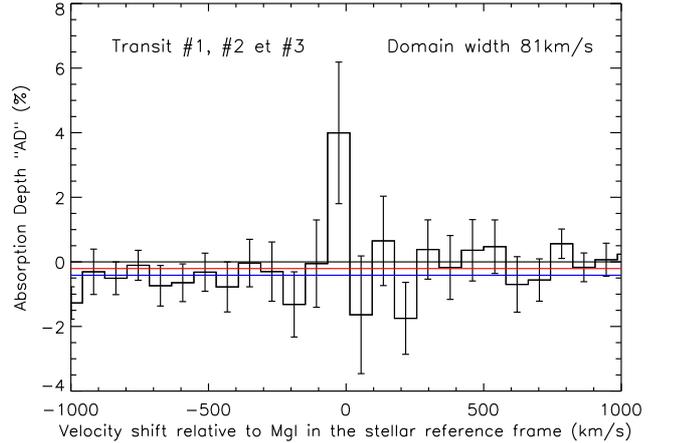} 
   \caption{Plot of the Mg\,{\sc i} $AD_{In/Out,{Tail}}$ 
   differences between the orbit~\#5 (before planetary transit) 
   and the orbit~\#2 (after transit) using a 81~km/s band width (histogram). 
   For comparison, the same differences have been measured in the stellar continuum 
   blue-ward and red-ward of the Mg\,{\sc i} line (blue and red lines, respectively).
  No differential absorption is detected, except in a single pixel that corresponds to a  
  blue-shifted velocity domain from $-67$ to $+14$\,km/s (see text).
}
   \label{post_transit}
\end{figure}

As expected, the evaluated differences are close to zero (Fig.~\ref{post_transit}). 
The average of the evaluations made over two symmetric broad spectral domains 
on both sides of the Mg\,{\sc i} line (over the -2\,500 to -500 km/s blue domain 
and the +500 to +2\,500 km/s red domain) is also nearly equal to zero.
However an outliers measurement near the Mg\,{\sc i} line core is obtained 
at a maximum differential 
$AD_{In/Out,{Tail}}$ of about 4.0$\pm$2.2\%\ over a blue-shifted, 81~km/s wide domain 
from $-67$ to $+14$~km/s in the stellar reference frame. 
This absorption signature, as detected in data collected after the planetary transit, 
suggests the presence of a tail of magnesium transiting after the planet.

We also find that the average of the $AD_{In/Out,{Tail}}$ differences 
estimated in the entire -2\,500 to +2\,500~km/s spectral range 
(ignoring the 4\%\ outlier) is equal to -0.4$\pm$0.4\%.  
This shows that the stellar flux variations on the average over the three transits 
are negligible and that stellar activity and star spots are not a concern in our study,
as previously discussed in Sect.~\ref{sec_visit_long_flux_corr}.

\begin{figure}[htb]
  \centering
    \includegraphics[width=\columnwidth]{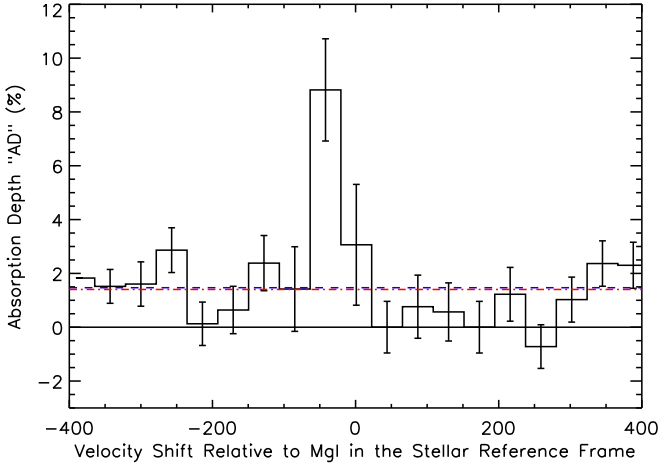}
   \caption{Plot of the absorption depth in 43~km/s band widths (histogram), as
   measured around the Mg\,{\sc i} line by using only the orbit~\#2 as the 
   out-of-transit reference. 
   The absorption depths measured in the stellar continuum 
   in the +500 to +2\,500~km/s range (red side continuum, red dot-dashed line) 
   and in the -2\,500 to -500~km/s range (blue side continuum, blue dashed line) are also shown. 
   The $AD'^{Total}$ absorption depth is measured to be 8.8$\pm$2.1\%
   in the velocity domain from $-67$ to $-19$\,km/s.}
   \label{MgI_43kms_Orb2}
\end{figure}

Because of this possible Mg\,{\sc i} absorption in orbit~\#5, the estimates of the transit absorption that are 
obtained using a fit with a model of an opaque spherical body could be biased.  
We thus repeated the fitting process with a modified de-trending model, in which only 
orbit~\#2 is used as the out-of-transit reference. 
We note the corresponding absorption depths $AD'$ (Fig.~\ref{MgI_43kms_Orb2}).
We obtain the following results~: 
$AD'^{Cont.}$=1.3$\pm$0.1\%, $AD'^{Total}$=8.8$\pm$2.1\% (4.2$\sigma$), 
and $AD'^{Excess}$=7.5$\pm$2.1\%.

Using a planetary radius given by (R$_P$/R$_*$)$^2$=1.444\%, 
an absorption of 8.8\%\ corresponds to about 2.5~R$_P$.
However, the Mg\,{\sc i} absorbing cloud of gas is likely asymmetric because the Mg\,{\sc i} absorption is detected only in the blue wing of the line 
(between -62 and -19~km/s);
this possibility is strengthened by the detection of a tail transiting after the planetary body. 
Therefore, these values of Mg\,{\sc i} altitude represent a lower limit and the detected 
Mg\,{\sc i} atoms can be much higher in the planetary atmosphere.

\subsection{Dynamics of escaping magnesium atoms}
\label{Dynamics of escaping magnesium atoms}

\begin{figure}[htb]
  \centering
  \includegraphics[height=\columnwidth,angle=-90]{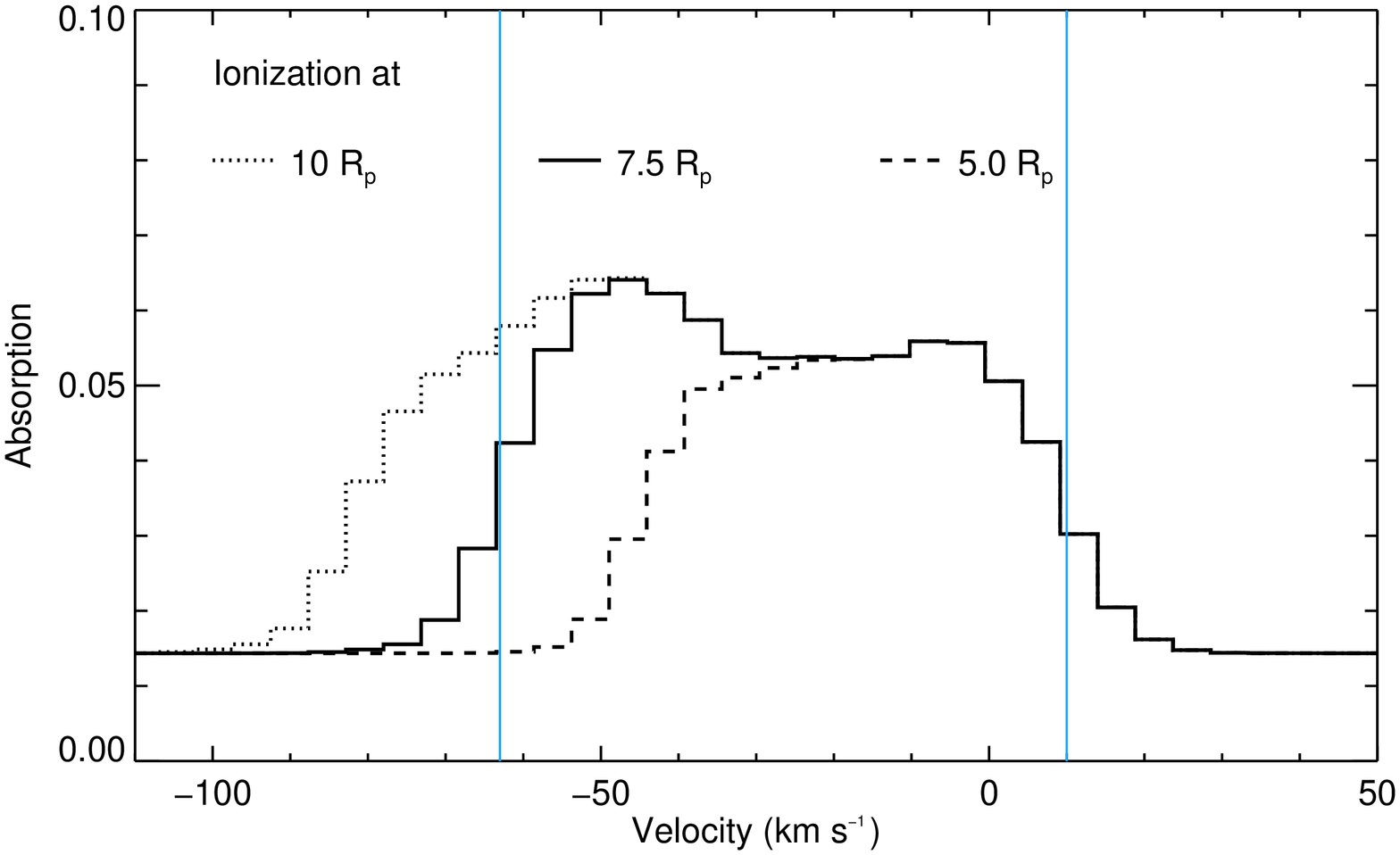}
  \includegraphics[height=\columnwidth,angle=-90]{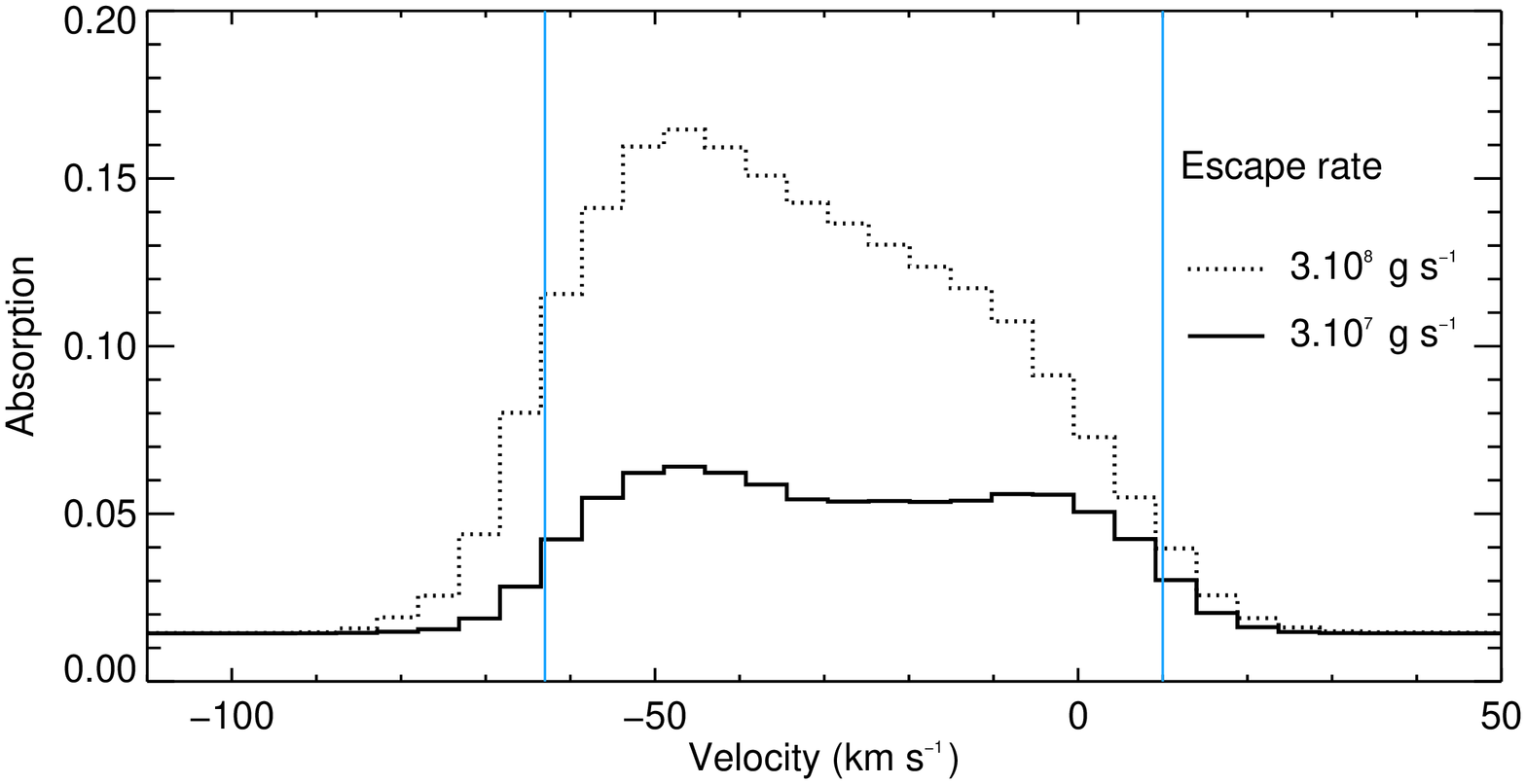} 
   \caption[]{Theoretical absorption depth of Mg\,{\sc i}, as calculated using numerical 
   simulations of the escape of magnesium atoms accelerated 
   by stellar radiation pressure.\\
   {\bf Upper panel.} 
   Plot of the $AD$ as a function of the radial velocity
   for various ionization cut-off radii (in R$_P$ units). 
   We find that a cut-off of 7.5~R$_P$ provides an $AD$ profile that matches
   the velocity range of the observed absorption on the blue edge 
   (indicated by the leftward vertical line); the other velocity limit shown is around 10~km/s
   (the other vertical line), which is
   independent of the radiation pressure model calculation, 
   since it is related to the assumed blow-off  
   velocity; this profile is obtained for 
   a magnesium mass loss rate of 3$\times 10^7$ g.s$^{-1}$.\\   
   {\bf Lower panel.} Same as above for an ionization cut-off radius of 
   7.5 R$_P$ and two different Mg\,{\sc i} escape rates. 
   This shows that the two main parameters of the model (escape 
   rate and cut-off radius) can be constrained independently through 
   the measurement of the absorption depth and the velocity range of the 
   gas. 
    }
   \label{simulation}
\end{figure}

All theoretical models with hydrodynamical escape ({\it e.g.} Koskinen et al.\ 2013b 
and references there in) predict the blow-off escape velocity to be in the 1 to 
10~km/s range (at $\sim$2$R_P$). Here, atomic magnesium is observed to be accelerated
after release, reaching velocities in the range of -62 to -19 km/s.
In the exosphere, Mg\,{\sc i} atoms are in a collisionless region and naturally 
accelerate away from the star by radiation pressure. 
As a result, magnesium atoms in the exosphere are expected 
to be at blue shifted velocities, as observed. 

We have developed a numerical simulation, following those done for H\,{\sc i} atoms 
(see Lecavelier des Etangs et al.\ 2012 for
HD\,189733\,b; Bourrier \& Lecavelier des Etangs 2013 for HD\,189733\,b and HD\,209458\,b) 
to calculate the dynamics of magnesium atoms. 
In the simulation, the Mg\,{\sc i} particles are released 
from a level R$_{0}$ within the atmosphere at the blow-off flow velocity, which is 
assumed to be 10~km/s. The atoms' dynamics are calculated by considering 
the gravity of the planet and the star, as well as the 
stellar radiation pressure. This last force acts in the opposite direction
of the stellar gravity and progressively accelerates 
the particles away from the star up to velocities similar to  
the observed blue shifted atmospheric absorption. We also consider
that radiation pressure is directly proportional to the 
velocity-dependent stellar flux, as seen by the moving particles. Because of 
the profile of the stellar line near 
the Mg\,{\sc i} resonance line (Fig.~\ref{MgI_spetrum}) 
with a stronger flux away from the line center, 
the particle acceleration increases with their velocity blueward of the line. 
The Mg\,{\sc i} atoms are ionized by the stellar flux but also recombined if the ambient 
electron volume density is high enough (see below). We represented 
the balance of these two effects by setting an ionization radius, 
R$_{\rm ion}$, above which magnesium atoms are all ionized. 
Above this cut-off radius, the electron volume density is too low 
to allow efficient recombination of Mg\,{\sc ii} into Mg\,{\sc i}. 
The third parameter of the simulation is the Mg\,{\sc i} escape rate, $\dot{M}[Mg]$.

The results of the numerical simulation are shown in Fig.~\ref{simulation} 
with two varying parameters: R$_{\rm ion}$ and dM/dt[Mg]. 
The parameter R$_{0}$ was fixed at 2~R$_P$ at about
the exobase level, where collisions become negligible (note that 
the value of R$_{0}$ is found to have no significant impact on the results). 
Our first aim was 
to reproduce the sharp velocity limit of the detected absorption signature 
around $-60$~km/s in the stellar reference frame. 
Our second goal was to reproduce 
the absorption depth measured in the Mg\,{\sc i} line 
as a function of the radial velocity, especialy with
the absorption of the velocity range [-62;-19 km/s], which is measured to be 
about 6\%. We found that the blueward velocity limit and the absorption depth
can be fitted in a straightforward manner, because they are constrained 
independently by the cut-off radius and the escape rate, respectively 
(see Fig.~\ref{simulation}). We found that our observations are well reproduced 
for R$_{\rm ion}$$\sim$$7.5$~R$_P$ and dM/dt[Mg]~$\sim$$3\times$$10^7$~g.s$^{-1}$. 
Assuming a solar abundance of magnesium ($\sim$$4\times$$10^{-5}$ with respect to hydrogen, 
see Asplund et al.\ 2009), this last result corresponds to an H\,{\sc i}
escape rate from HD\,209458\,b of $\sim$3$\times$$10^{10}$~g.s$^{-1}$. This is consistent 
with the escape rate of neutral hydrogen as derived from Lyman-$\alpha$ 
observations (see {\it e.g}, Lecavelier des Etangs et al.\ 2004; Ehrenreich et al.\ 2008; Bourrier \& Lecavelier des Etangs\ 2013). 

The planetary radius that corresponds to the $AD$ of 6--8\%\ ($\sim$2.5\,$R_P$) is different 
from the estimated cut-off radius R$_{ion}$ of Mg\,{\sc i} at about 7.5~R$_P$.
This is explained because the 2.5\,$R_P$ estimation assumes 
a cloud with a spherical symmetry, while the
geometry of the occulting gas is not spherically symmetric 
but pushed by radiation pressure in a comet-like tail that trails behind the planet. 
Of course, this requires a cloud that is more extended on only one side to produce 
a similar absorption depth. 

\subsection{Magnesium ionization and recombination}
 
The mere presence of atomic magnesium may appear surprising in the exosphere
of HD\,209458\,b, because this neutral species has a low ionization potential of 7.65~eV.
As a comparison to the HD\,209458\,b environment, carbon is found in its ionized form
(Vidal-Madjar et al.\ 2004; Linsky et al.\ 2010), while the carbon ionization
potential (11.26~eV) is larger than that for magnesium (7.65~eV).

However, there is a similar situation in the local interstellar medium,
where neutral atomic magnesium is observed. To explain such a paradoxical
situation, it was shown that the Mg\,{\sc ii} recombination is extremely sensitive
to the medium temperature (see Fig.~8 of Lallement et al.\ 1994).
Indeed the dielectronic recombination rates of Mg\,{\sc ii} steeply rises at temperatures
above 6\,000~K (Jacobs et al.\ 1979). With this mechanism being the most effective
to recombine Mg\,{\sc ii} at temperature above $\sim$3\,000\,K,
the balance between that recombination mechanism
and all ionization mechanisms is such that Mg\,{\sc i} can survive in this 
astrophysical environment.

\begin{figure}
   \includegraphics[height=\columnwidth,angle=+90]{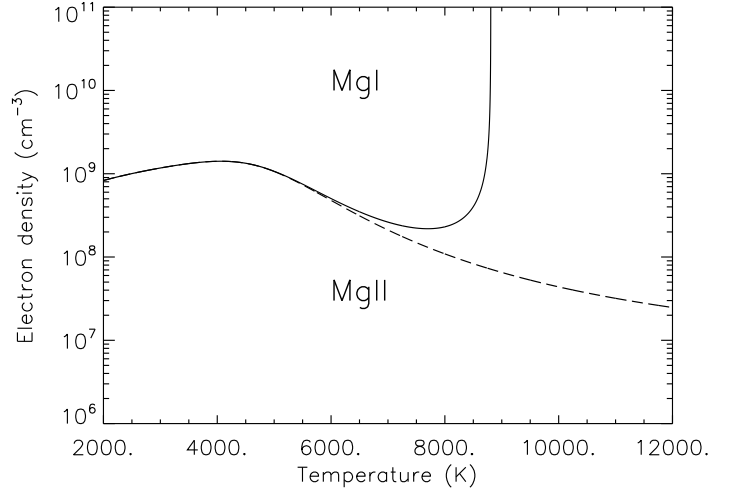}
\caption{
Plot of the electron volume density needed to obtain the magnesium ionization
equilibrium as a function of the temperature. At a lower density, the magnesium
is mostly ionized, while the magnesium is mostly in its atomic form at higher density.
The ionization rate is calculated using
the electron-impact ionization rate extrapolated from the rate
of Voronov (1997) that is combined with the photo-ionization rate (solid line)
or the photo-ionization rate only (dashed line).
}
\label{Fig: ne_vs_T}
\end{figure}

In the planetary atmosphere, the detection of atomic
magnesium through the Mg\,{\sc i} line shows that
the electronic volume density must be large enough to allow the Mg\,{\sc ii} recombination
to overcome all the possible ionization mechanisms.
These ionization mechanisms are~: i) the UV photo-ionization from the nearby star,
ii) the  electron-impact ionization (see {\it e.g.} Voronov 1997),
iii) the charge-exchange of Mg\,{\sc i} with protons (see {\it e.g.} Kingdon \& Ferland 1996)
and iv) the charge-exchange of Mg\,{\sc i} with He\,{\sc ii}.

Because the situation is certainly far from equilibrium in these upper parts
of the atmosphere, the electron and proton volume densities are likely different.
He\,{\sc ii} may also play a non-negligible role.
It is thus very difficult to evaluate the charge-exchange ionization rates.
The electron-impact ionization may also be significant at very high temperatures.
For instance, the ionization overcomes the dielectronic recombination
at temperature above $\sim$9\,000\,K by using extrapolation of electron-impact ionization rates given by
Voronov (1997).
However, the rates given by Voronov (1997) are given for temperatures
above $\sim$11\,000\,K and may not apply at lower temperature regimes.
Because of the large uncertainties on the effective rate of electronic ionization
at temperatures below 11\,000\,K, we decided to consider both situations with
and without this ionization mechanism (Fig.~\ref{Fig: ne_vs_T}).

Among the ionization mechanisms, the photoionization can be easily calculated.
To estimate ionizing UV flux at 0.047~AU from HD\,209458,
we used $F_{\rm HD\,209458\,b}(\lambda)$, the stellar
UV flux at wavelengths below the ionization threshold of 1621~\AA\ as measured by
Vidal-Madjar et al.\ (2004). The photoionization cross-section as a function of
wavelength $\sigma_{\rm ion}(\lambda)$ for the $3s^2\ ^1S$ ground state of Mg\,{\sc i}
is taken from Merle et al.\ (2011). We thus found the magnesium ionization rate to be
$$
\Gamma_{\rm ion}=\int^{1621\AA} \frac{F_{\rm HD\,209458\,b}(\lambda)\sigma_{\rm ion}(\lambda)}{hc}
                 \lambda d\lambda = 4.6\times 10^{-4}  s^{-1}.
$$
At the short orbital distance from the star, the magnesium is therefore
quickly photoionized with a short timescale of about 0.6~hours.
The presence of atomic magnesium at altitudes up to several planetary radii
from the planet can hence be explained only if the recombination is efficient
enough to compensate for the ionization.

The radiative and dielectronic recombination rates can be found in the literature. 
These recombination rates depend upon two atmospheric parameters:
the electron temperature and the electron volume density ({\it e.g.}, Frisch et al.\ 1990).
The dielectronic recombination rate, $\alpha_{\rm rec\,di}$,
is highly temperature-dependent and quickly rises at temperatures above 6\,000~K.
We used the dielectronic rate given by
$\alpha_{\rm rec\, di} =1.7\times 10^{-3} T^{-1.5} \exp(-T^0/T)$\,cm$^3$s$^{-1}$
with $T^0=5.1\times 10^4$\,K 
and the radiative rate given by
$\alpha_{\rm rec\, rad} =1.4\times 10^{-13} (T/10^4)^{-0.855}$\,cm$^3$s$^{-1}$
(Aldrovandi \& Pequignot 1973).
The total recombination rate is given by
$\alpha_{\rm rec}$=$\alpha_{\rm rec\, di}$+$\alpha_{\rm rec\, rad}$.

To estimate the electronic temperature and volume density needed to explain the
detection of atomic magnesium, we calculated the
ionization-recombination balance of magnesium exposed to the ionizing
UV flux at 0.047~AU from HD\,209458.
Because of the unknown efficiency of ionization mechanisms
other than photoionization,
we decided to consider only photoionization and possibly the electron-impact
ionization rate extrapolated from Voronov (1997).
With this limitation, the calculated ionization rate is a lower limit, and
the corresponding electron density is also a lower limit.
This lower limit on the electronic density that is needed to obtain the
magnesium ionization equilibrium is calculated as a function of the temperature
(Fig.~\ref{Fig: ne_vs_T}).
It is noteworthy that the required electronic density must be above $10^8$--$10^9$\,cm$^{-3}$,
which is in reasonable agreement with the value estimated
by Koskinen et al.\ (2013a; see their Fig.~9). This value is simply deduced from the 
Koskinen et al.\ (2013a) model proton
density by assuming that the medium is neutral, which should be true somewhere below the altitudes considered here.

If the rise in electron-impact ionization efficiency, as given by Voronov (1997) applies
to this temperature regime, we conclude that the temperature must be below $\sim$9\,000\,K
(Fig.~\ref{Fig: ne_vs_T}). The larger efficiency of the dielectroniic recombination also
favors temperature above $\sim$6\,000\,K.
Thus, the present detection of magnesium provides new constraints
on the density and temperature, high in the exosphere of HD\,209458\,b,
up to $\sim$7.5~$R_P$.
Previously, the measurement of the temperature at the highest altitude
in the thermosphere was obtained through analysis of the
absorption in the core of the NaI line (Vidal-Madjar et al.\ 2011a, 2011b); this
constrained a temperature of $\sim$3\,600\,K at $10^{-7}$\,bar, which corresponded
to a density of $\sim$$2\times 10^{11}$\,cm$^{-3}$.
Here, the detection of atomic magnesium, which requires high temperature and electron
density at high altitudes in the atmosphere,
extends the temperature-altitude profile in the thermosphere, which is 
already probed deeper through the sodium line profile. 
With a lower limit for the electron density of
$10^8$--$10^9$\,cm$^{-3}$ typical of the base of the exosphere
and velocity characteristics of acceleration by radiation pressure,
the detected magnesium atoms provide a new insight into
the transition region between the upper part of the thermosphere
and the bottom of the exosphere, where the atmospheric escape is taking place.     
  
\section{Conclusion}

Here, we presented the detection of absorption in the Mg\,{\sc i} line 
during the planetary transit, which is due to magnesium atoms at high altitude 
in the atmosphere of HD\,209458\,b. 
We measured an atmospheric absorption of 6.2$\pm$2.9\% in the velocity range 
from -62 to -19~km/s, which presents a false-positive probability of being 
due to noise of about 6\%. 
In addition, no extra absorptions in the Mg\,{\sc ii} lines were found 
with an upper limit of about one percent. 

From this Mg\,{\sc i} detection, we derived the following consequences:

-- The high ionization rate of magnesium by the stellar UV photons implies that
dielectronic recombination of Mg\,{\sc ii} must be efficient, which requires a high electronic
density above $10^8$--$10^9$\,cm$^{-3}$.

-- Because thermal escape of heavy species like magnesium is inefficient (even at 10\,000~K),
the detection of escaping magnesium confirms the scenario of an hydrodynamical escape 
(blow-off scenario).
 
-- A numerical simulation shows that the velocity range of observed magnesium
is easily 
explained through acceleration by radiation pressure. Pushed away by radiation 
pressure, the magnesium must be 
in atomic form for distances up to about 7.5 planetary radii for the planet center.

-- This numerical simulation also allows an estimate of the magnesium escape rate 
of $\sim$$3\times$$10^7$~g/s, which is consistent with the measured H\,{\sc i} 
escape rate, if we assume a solar abundance. 

-- Finally, a cometary tail geometry of the magnesium, as predicted by the numerical simulation, 
is possibly detected through residual absorption in the post-transit data. 
If true, the estimate of the absorption depth of Mg\,{\sc i} must be revised to a higher value 
of about 8.8$\pm$2.1\%.

\begin{acknowledgements}

Based on observations made with the NASA/ESA Hubble
Space Telescope, obtained at the Space Telescope Science Institute, which is
operated by the Association of Universities for Research in Astronomy, Inc.,
under NASA contract NAS 5-26555. These
observations are associated with program \#11576. 
The authors acknowledge financial
support from the Centre National d'\'Etudes Spatiales (CNES).
The authors also acknowledge the support of the French Agence Nationale de la Recherche (ANR), under program ANR-12-BS05-0012 "Exo-Atmos". 
J.-M.D. acknowledges funding from NASA through the Sagan Exoplanet Fellowship program administered by the NASA Exoplanet Science Institute (NExScI).
D.K.S. and C. M. H. acknowledges support from STFC consolidated grant ST/J0016/1.
D.E. acknowledges the funding from the European Commission's Seventh Framework Programme as a Marie Curie Intra-European Fellow (PIEF-GA-2011-298916).

J. C. McConnell passed away on July 29, 2013. He was a highly respected scientist.
We all would like to dedicate the present study to him. John ``Jack'' 
was more than a colleague, a friend; we deeply miss him. Jack
was the first who understood that the atmospheric blow-off 
mechanism, which had never been directly observed before,
as the phenomenon, which takes place in the HD\,209458\,b atmosphere.

\end{acknowledgements}

\end{document}